\documentclass[twocolappendix,numberedappendix,iop]{emulateapj}
\usepackage{apjfonts}
\usepackage{subfigure}
\usepackage{color}
\usepackage{mathtools}
\usepackage[breaklinks,colorlinks,
  urlcolor=blue,citecolor=blue,linkcolor=blue]{hyperref}
\usepackage[all]{hypcap}

\DeclareRobustCommand{\orderof}{\ensuremath{\mathcal{O}}}
\newcommand{\new}[1]{{#1}}

\slugcomment{Draft version May 17, 2016}

\shorttitle{Progenitor-dependent Explosion Dynamics in Axisymmetric CCSN Simulations}
\shortauthors{Summa et al.}

\begin{document}


\title{Progenitor-dependent Explosion Dynamics in Self-consistent, Axisymmetric Simulations\\of Neutrino-driven Core-collapse Supernovae}


\author{Alexander Summa\altaffilmark{1}, Florian Hanke\altaffilmark{1,2}, Hans-Thomas Janka\altaffilmark{1}, Tobias Melson\altaffilmark{1,2}, \\Andreas Marek\altaffilmark{3} and Bernhard M\"uller\altaffilmark{4,5}}

\email{asumma@mpa-garching.mpg.de}
\email{thj@mpa-garching.mpg.de}
\altaffiltext{1}{Max-Planck-Institut f\"ur Astrophysik, Karl-Schwarzschild-Str.~1, D-85748 Garching, Germany}
\altaffiltext{2}{Physik Department, Technische Universit\"at M\"unchen, James-Franck-Str.~1, D-85748 Garching, Germany}
\altaffiltext{3}{Max Planck Computing and Data Facility (MPCDF), Gie{\ss}enbachstr.~2, D-85748 Garching, Germany}
\altaffiltext{4}{Astrophysics Research Centre, School of Mathematics and Physics, Queen's University Belfast, Belfast, BT7~1NN, United Kingdom}
\altaffiltext{5}{Monash Centre for Astrophysics, School of Physics and Astronomy, Monash University, Victoria 3800, Australia}


\begin{abstract}
We present self-consistent, axisymmetric core-collapse supernova 
simulations performed with the \textsc{Prometheus-Vertex} code for 
18 pre-supernova models in the range of 11--28\,$\mathrm{M}_\odot$,
including progenitors recently investigated by other groups.
All models develop explosions, but depending on the progenitor 
structure, they can be divided into two classes. With a steep density
decline at the Si/Si-O interface, the arrival of this interface at the
shock front leads to a sudden drop of the mass-accretion rate, triggering
a rapid approach to explosion. With a 
more gradually decreasing accretion rate, it takes longer for the 
neutrino heating to overcome the accretion ram pressure and explosions set
in later. Early explosions are facilitated by high mass-accretion rates
after bounce and correspondingly high neutrino luminosities combined 
with a pronounced drop of the accretion rate and ram pressure at the 
Si/Si-O interface. Because of rapidly shrinking neutron star radii and
receding shock fronts after the passage through their maxima, our models exhibit
short advection time scales, which favor the efficient growth of the standing
accretion-shock instability. The latter plays a supportive role at 
least for the initiation of the re-expansion of the stalled shock before runaway.
Taking into account the effects of turbulent pressure in the gain layer,
we derive a generalized condition for the critical neutrino luminosity
that captures the explosion behavior of all models very well. We validate
the robustness of our findings by testing the influence of stochasticity,
numerical resolution, and approximations in some aspects of the microphysics.
\end{abstract}





\section{Introduction}

Nearly half a century after the first suggestion \citep{Colgate1966} that 
neutrinos might play an important role in core-collapse supernovae (CCSNe), the 
viability of the delayed neutrino-driven mechanism \citep{Bethe1985} is still 
controversially discussed. Although the degree of sophistication of the 
explosion models has continuously increased and a growing number of 
multidimensional simulations have been conducted over the past years, the 
conclusions with respect to the neutrino-driven mechanism are contradictive and 
an unambiguous verification of the physics that drives the explosion has not yet 
been possible.

While successful explosions with simulations in spherical symmetry (1D) 
including state-of-the-art physics could only be obtained in cases of stars with 
O-Ne-Mg and low-mass Fe cores 
\citep{Kitaura2006,Janka2008,Fischer2010,Melson2015}, explosion models in two 
dimensions (i.e.\,with assumed axisymmetry; 2D) demonstrated the important and supportive role of 
multidimensional effects. However, the 2D results reported by various groups 
differ considerably. According to the results by, for example, 
\citet{Marek2009a}, \citet{Janka2012}, \citet{Suwa2010,Suwa2016}, \citet{Nakamura2015}, \citet{Mueller2012,Mueller2012c}, and \citet{Mueller2014}, 
simulations in axisymmetry show rather late explosions with energies seemingly below the 
canonical value of $10^{51}\,\mathrm{erg}$ for typical CCSNe. Of course, it has 
to be noted that not all simulations were continued to the time when a 
saturation of the explosion energy can be expected. 
\citet{Bruenn2013,Bruenn2016} presented four 2D simulations for progenitors with 
zero-age main sequence (ZAMS) masses between 12\,M$_\odot$ and 25\,M$_\odot$ where the explosions already 
begin at fairly early times after bounce ($\sim0.2$\,s) and the explosion energies are in reach of 
those deduced from observations. Curiously, in spite of the different structures of the 
four progenitor models, all explosions (i.e.\,runaway shock expansions) set in nearly at the same time. Using the 
same four progenitor models, but different treatments concerning hydrodynamics, gravity,  
equation of state (EoS), and neutrino transport, \citet{Dolence2015} did not find any 
explosion, 
while \citet{Skinner2015} and \citet{OConnor2015} reported failures or successes that depended
on the applied gravity (Newtonian or relativistic potential) and transport treatment \citep[for a summary of
recent 2D results, the reader is also referred to][]{Janka2016}.
This unsatisfactory situation clearly underlines the need for more 
detailed tests and code comparisons among the different CCSN simulation groups 
in the future. 

The imposed symmetry constraints in 2D simulations are also the cause of 
drawbacks. The unipolar or bipolar deformations along the symmetry axis observed 
in 2D models seem to be strongly connected to the artificial assumption of 
rotational symmetry, and the inverse turbulent energy cascade distributes the 
energy in an unphysical way to the largest scales \citep[see][]{Kraichnan1967,Hanke2012,Couch2013,Radice2016}. 
But due to the huge 
computational demands of self-consistent simulations in three dimensions
\citep[see e.g.][]{Hanke2013a,Tamborra2013a,Tamborra2014,Tamborra2014a,Takiwaki2014,Melson2015,Melson2015a,Mueller2015a,Lentz2015,Kuroda2012,Kuroda2016}, 
systematic studies of larger sets of progenitor models or detailed 
investigations of different explosion parameters are restricted to the 
axisymmetric modeling approach at the moment. Even in 2D, investigations
of a wider range of pre-supernova models usually employ only
simplified neutrino transport schemes \citep[e.g.,][]{Nakamura2015,Pan2016,Suwa2016}.

In the following, we report the results of 2D simulations with the 
\textsc{Prometheus-Vertex} code from \citet{Hanke2014}. The consideration of a large set of 18 different 
pre-supernova models allows us to investigate the influence of the progenitor 
structure on the explosion physics in a systematic way, and a connection of 
progenitor properties to certain aspects of the evolution of the supernova 
explosion becomes possible. Besides a set of 14 pre-supernova models from 
\citet{Woosley2002}, we include the four progenitors from \citet{Woosley2007} that 
were chosen by  \citet{Bruenn2013,Bruenn2016}, \citet{Dolence2015}, \citet{OConnor2015}, and \citet{Skinner2015} and 
discuss our simulation results of these four models in depth. This is intended 
to facilitate future comparisons between the different simulation groups and will 
hopefully help to shed light on the currently rather diffuse situation regarding the 
outcomes of CCSN simulations with different codes.

Motivated by the question why the explosions in our models set in at largely different 
times without any obvious connection to special values of individual parameters like the 
non-radial kinetic energy, heating efficiency or maximum/average entropy in the 
gain layer, we will also present a theoretical analysis that sets our results into
the context of the critical luminosity concept for the initiation of neutrino-driven
explosions. We will show that the critical condition of $L_\nu\langle E_\nu^2\rangle$
as a function of $\dot{M}M_\mathrm{NS}$ coined by \citet{Mueller2015} ($L_\nu$ denotes the total electron-flavor neutrino luminosity,
$\langle E_\nu^2\rangle$ the weighted average of the mean squared energies of 
electron neutrinos and antineutrinos, $\dot{M}$ the mass-accretion rate, and $M_\mathrm{NS}$ the 
mass of the proto-neutron star, see also Sect.\,\ref{Crit}) defines a universal relation
that yields an excellent description of the behavior of our models at the transition
to explosion, provided the effects of turbulent pressure as well as corrections due to the
time- and model-dependent variations of the gain radius and binding energy in the gain layer are taken into account.

The paper is structured as follows. After a brief summary of the numerical 
setup in Sect.\,\ref{num_set}, our simulation results are presented in 
Sect.\,\ref{res_dis}. In Sect.\,\ref{Crit}, we show that the approach to explosions
of our model set can be well described by a generalized version of the critical luminosity
condition. We conclude in Sect.\,\ref{concl} and close the paper
with appendices where detailed information for some special aspects is provided and the 
dependence of our results on numerical
resolution and stochastic effects is discussed. We also briefly
describe the influence of special microphysics (in particular neutrino pair-conversion 
and $\nu-\nu$ scattering processes as well as nucleon correlations and 
reduced effective nucleon masses at high densities), which 
are not included by other groups \citep[e.g.][]{Bruenn2013,Bruenn2016}.

\section{Numerical setup}\label{num_set}

All calculations presented in this paper were performed with the
elaborate neutrino-hydrodynamics code 
\textsc{Prometheus-Vertex}. This tool for the simulation of CCSNe couples the 
hydrodynamics solver \textsc{Prometheus} 
\citep{Fryxell1989} via lepton number, energy, and momentum source terms with 
the neutrino transport module \textsc{Vertex} \citep{Rampp2002b}. The 
hydrodynamics module is based on a dimensionally split, time-explicit 
implementation of the Piecewise Parabolic Method of \citet{Colella1984}, which 
is a conservative, Godunov-type scheme with higher-order spatial and temporal 
accuracy that employs an exact Riemann solver. The transport module 
\textsc{Vertex} is a time-implicit solver for the energy- and velocity-dependent 
0\textsuperscript{th} and 1\textsuperscript{st} order moment equations for 
neutrinos and antineutrinos of all flavors. The system of moment equations is 
closed by a variable Eddington factor obtained by solving model Boltzmann 
equations iteratively up to convergence on all angular grid bins, called 
``radial rays''. This ``ray-by-ray'' approximation implies that the neutrino 
radiation field is assumed to be axially symmetric around the radial direction at each spatial point. 
Non-radial components of the neutrino flux  are thus ignored except for explicitly included terms 
associated with non-radial neutrino-pressure gradients and non-radial advection of 
the neutrinos when trapped in the stellar fluid \citep[``ray-by-ray-plus 
approach'', cf.][]{Buras2006a}. The energy dependence of the 
transport is fully retained. Gravitational redshifting, all velocity-dependent 
$\orderof\left( v/c\right)$ terms like Doppler shifts, and the redistribution of 
neutrinos in energy space by non-isoenergetic scatterings of all types of 
targets (nucleons, electrons, neutrinos) are included with the most sophisticated 
treatment of neutrino interactions presently available \citep[see, 
e.g.,][]{Marek2009a,Mueller2012c}. For more details about the 
\textsc{Prometheus-Vertex} code and the 
applied numerics, the reader is referred to \citet{Rampp2002b}, \citet{Buras2006a}.

The simulations were conducted with a 2D gravitational potential 
\citep[cf.][]{Buras2006a} including general relativistic monopole corrections as 
described in \citet{Marek2006a}. At high densities, the EoS of 
\citet{Lattimer1991} with a nuclear incompressibility of 220\,MeV and a symmetry 
energy parameter of 29.3\,MeV was used. Below a certain density and above
a certain temperature, which were chosen differently before and after bounce,
we applied a low-density EoS for nuclear statistical equilibrium (NSE) 
with 23 nuclear species. Below NSE temperature (chosen to be 0.5\,MeV in the
present simulations) we apply the flashing treatment of \citet{Rampp2002b} as an
approximate description of nuclear burning.
The axisymmetric models were computed on 
a spherical polar grid with initially 400 radial and 128 angular zones. The 
radial zones were non-equidistantly distributed from the center with a reflecting 
boundary condition at the coordinate origin to an outer boundary of $10^9$\,cm with an inflow condition. 
During the simulations, the radial grid was gradually refined to ensure adequate 
resolution in the proto-neutron star surface region. At the time the simulations were
stopped, the number of radial grid zones typically amounted to $\sim 600$, and a resolution
of $\Delta r/r \sim 3.5\times 10^{-3}$ at the proto-neutron star surface was reached.
Tests with higher resolution in radial and angular directions will also be
presented in Appendix \ref{res_dep}. The innermost 1.6\,km of 
the stellar core (corresponding to the innermost six radial zones) were 
treated in spherical symmetry to avoid excessive time step limitations at the 
center of the spherical grid. At 10\,ms after core bounce, seed perturbations of 
0.1\,\% in density were randomly introduced on the entire computational domain in order to trigger the
growth of aspherical instabilities in the previously spherically symmetric stellar 
progenitor models. For the neutrino transport, 12 geometrically spaced energy bins 
with an upper bound of 380\,MeV were employed.

\begin{figure*}
\flushleft
\includegraphics[width=\textwidth]{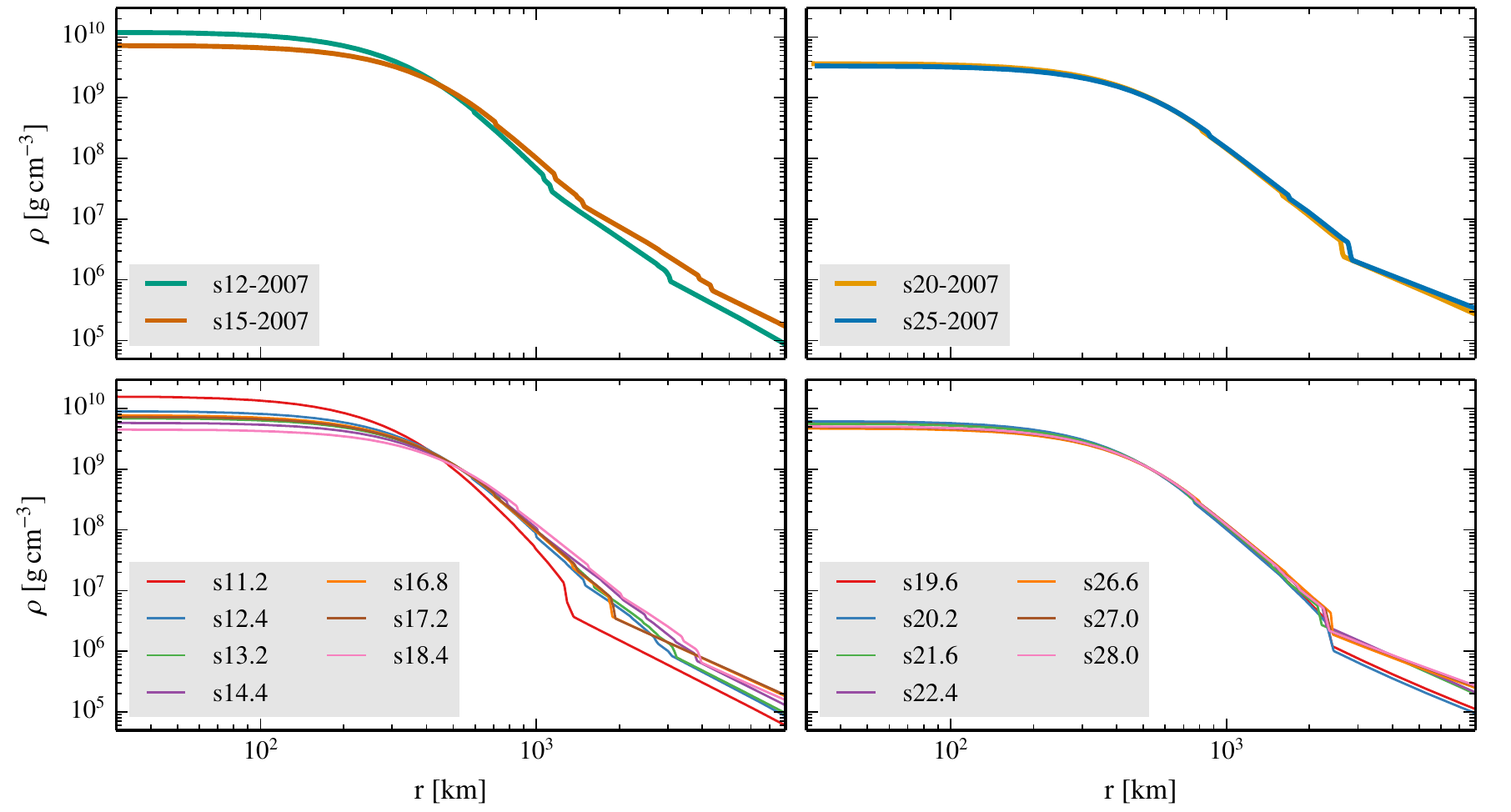}
\caption{Radial density profiles for the 18 progenitors of \citet{Woosley2007} 
(first row) and \citet{Woosley2002} (second row) at the onset of core 
collapse. The models with lower ZAMS masses are displayed in the left column, 
models with higher ZAMS masses in the right column.\label{denplot}}
\end{figure*}

\setlength\tabcolsep{4pt}
\capstartfalse
\begin{deluxetable*}{@{}c@{}ccccccccccccc}

\tabletypesize{\scriptsize}
\centering
\tablecaption{Overview of characteristic properties of the simulated axisymmetric models\label{tab1}}

\tablehead{&
\colhead{$M_\mathrm{ZAMS}$\tablenotemark{a}}\vspace{-0.1cm} & & & & \colhead{$t_\mathrm{exp}$\tablenotemark{c}} & 
\colhead{$R_\mathrm{s}(t_\mathrm{exp})$\tablenotemark{d}} & 
\colhead{$R_\mathrm{s}^\mathrm{m}(t_\mathrm{exp})$\tablenotemark{e}} & 
\colhead{$R_\mathrm{NS}(t_\mathrm{exp})$\tablenotemark{f}} & 
\colhead{$L_{\nu_\mathrm{e}}(t_\mathrm{exp})$\tablenotemark{g}} &
\colhead{$\dot{M}(t_\mathrm{exp})\tablenotemark{h}$} &
\colhead{$M_\mathrm{NS}(t_\mathrm{exp})$\tablenotemark{i}} &
\colhead{$t_\mathrm{exp}^*$\tablenotemark{j}} &
\colhead{$R_\mathrm{s}^\mathrm{m}(t_\mathrm{exp}^*)$\tablenotemark{k}}
\\
\colhead{model}\vspace{-0.1cm} & & 
\colhead{$\xi_{1.5}$\tablenotemark{b}} &
\colhead{$\xi_{1.75}$\tablenotemark{b}} & 
\colhead{$\xi_{2.5}$\tablenotemark{b}}
\\
& \colhead{[M$_\odot$]} & & & &
\colhead{[ms]} & 
\colhead{[km]} & 
\colhead{[km]} & 
\colhead{[km]} & 
\colhead{[$10^{52}\,\mathrm{erg}\,\mathrm{s}^{-1}$]} & 
\colhead{[M$_\odot\,\mathrm{s}^{-1}$]} & 
\colhead{[M$_\odot$]} & 
\colhead{[ms]} & 
\colhead{[km]}}

\startdata
\textit{Model Set I}\\
s12-2007 & 12.0 & 0.612 & 0.234 & 0.023 & 743 & 142 & 269 & 24 & 2.208 & 0.142 & 1.58 & 815 & 743 \\
s15-2007 & 15.0 & 0.878 & 0.547 & 0.182 & 550 & 119 & 208 & 27 & 3.660 & 0.395 & 1.77 & 631 & 686 \\
s20-2007 & 20.0 & 1.003 & 0.769 & 0.286 & 292 & 191 & 306 & 36 & 4.007 & 0.386 & 1.83 & 358 & 1015 \\
s25-2007 & 25.0 & 1.009 & 0.819 & 0.330 & 338 & 144 & 224 & 33 & 4.195 & 0.374 & 1.92 & 407 & 659 \\
\textit{Model Set II}\\
s11.2 & 11.2 & 0.194 & 0.073 & 0.005 & 332 & 251 & 519 & 34 & 1.845 & 0.113 & 1.33 & 349 & 731 \\
s12.4 & 12.4 & 0.759 & 0.265 & 0.028 & 634 & 172 & 317 & 25 & 2.176 & 0.142 & 1.61 & 697 & 426 \\
s13.2 & 13.2 & 0.821 & 0.335 & 0.049 & 597 & 161 & 298 & 26 & 2.298 & 0.152 & 1.66 & 664 & 885 \\
s14.4 & 14.4 & 0.868 & 0.515 & 0.124 & 726 & 120 & 201 & 24 & 2.852 & 0.198 & 1.79 & 799 & 513 \\
s16.8 & 16.8 & 0.821 & 0.355 & 0.159 & 472 & 173 & 310 & 29 & 2.536 & 0.246 & 1.59 & 543 & 600 \\
s17.2 & 17.2 & 0.857 & 0.367 & 0.168 & 382 & 179 & 289 & 32 & 2.886 & 0.263 & 1.58 & 453 & 751 \\
s18.4 & 18.4 & 0.955 & 0.652 & 0.188 & 520 & 118 & 198 & 28 & 3.866 & 0.346 & 1.85 & 583 & 606 \\ 
s19.6 & 19.6 & 0.873 & 0.298 & 0.119 & 356 & 206 & 369 & 33 & 2.354 & 0.145 & 1.61 & 415 & 699 \\
s20.2 & 20.2 & 0.840 & 0.249 & 0.106 & 346 & 194 & 328 & 33 & 2.480 & 0.125 & 1.59 & 414 & 718 \\
s21.6 & 21.6 & 0.939 & 0.467 & 0.181 & 503 & 169 & 275 & 28 & 2.772 & 0.266 & 1.70 & 572 & 744 \\
s22.4 & 22.4 & 0.960 & 0.527 & 0.200 & 393 & 161 & 267 & 32 & 3.233 & 0.291 & 1.71 & 459 & 674 \\
s26.6 & 26.6 & 0.960 & 0.569 & 0.228 & 326 & 228 & 363 & 34 & 2.938 & 0.249 & 1.71 & 373 & 677 \\
s27.0 & 27.0 & 0.960 & 0.524 & 0.233 & 389 & 208 & 314 & 32 & 2.918 & 0.263 & 1.71 & 453 & 650 \\
s28.0 & 28.0 & 0.962 & 0.524 & 0.236 & 400 & 157 & 256 & 32 & 3.240 & 0.258 & 1.71 & 474 & 833
\enddata
\tablenotetext{a}{ZAMS mass of the pre-supernova progenitor model.}
\tablenotetext{b}{Compactness parameter as defined in Eq.\,(\ref{eq:comp}) (calculated from the pre-supernova model).}
\tablenotetext{c}{Onset of explosion defined by the point in time when the ratio of advection to heating time scale reaches unity.}
\tablenotetext{d}{Mean shock radius at the onset of the explosion.}
\tablenotetext{e}{Maximum shock radius at the onset of the explosion.}
\tablenotetext{f}{Neutron star radius at the onset of the explosion (defined by the location of density $10^{11}\,\mathrm{g\,cm^{-3}}$).}
\tablenotetext{g}{Luminosity of electron neutrinos at the time of explosion (evaluated at 400\,km and given for an observer in the lab frame at infinity).}
\tablenotetext{h}{Mass-accretion rate at the onset of the explosion as defined in Eq.\,(\ref{eq:accr}) (evaluated at a radius of 400\,km).}
\tablenotetext{i}{Neutron star mass at the onset of the explosion (defined by the density surface of $10^{11}\,\mathrm{g\,cm^{-3}}$).}
\tablenotetext{j}{Point in time when the mean shock radius reaches a value of 400\,km.}
\tablenotetext{k}{Maximum shock radius at the point in time when the mean shock radius reaches a value of 400\,km.}
\end{deluxetable*}
\capstarttrue

\section{Results and Discussion}\label{res_dis}

This section is subdivided into two parts. First, we present the simulation 
results of four pre-supernova progenitor models from \citet{Woosley2007} in detail (Model Set I). 
The intention is to facilitate comparisons to recent publications of other groups 
that only focused on this set of progenitors \citep[e.g.][]{Bruenn2013,Bruenn2016,Dolence2015}. Our main findings 
regarding 14 pre-supernova models of \citet{Woosley2002} are discussed in the 
second part (Model Set II). The choice of these 14 models was guided by the 
results of the parametric study of \citet{Ugliano2012} with respect to promising 
candidates for successful explosions. Due to the fact that all 18 models explode 
within the framework of our self-consistent and physically highly elaborate 
simulations, the neutrino-driven explosion mechanism proves to be 
viable for a large set of progenitors with different 
ZAMS masses (at least in axisymmetry).  

An overview of all 18 explosion models and their characteristic properties 
is given in Table \ref{tab1}. Besides the ZAMS mass, 
$M_\mathrm{ZAMS}$, the compactness parameter defined by \citet{OConnor2011},
\begin{equation}
\xi_\mathrm{M} = 
\frac{M/\mathrm{M}_\odot}{R(M_\mathrm{bary}=M)/1000\,\mathrm{km}},
\label{eq:comp}
\end{equation}
is given for $M=1.5$\,M$_\odot$, 1.75\,M$_\odot$, and 2.5\,M$_\odot$ (calculated from the 
pre-supernova model). At the onset of the explosions (defined by the time 
$t_\mathrm{exp}$ when the ratio of advection and heating time scale reaches 
unity, see below), the mean shock radius, $R_\mathrm{s}$, the maximum shock radius, $R_\mathrm{s}^\mathrm{m}$,
the neutron star radius, $R_\mathrm{NS}$, the electron neutrino luminosity, $L_{\nu_\mathrm{e}}$, 
the mass-accretion rate, $\dot{M}$, and the (baryonic) neutron star mass, $M_\mathrm{NS}$ 
(defined as the matter with densities above $10^{11}\,\mathrm{g}\,\mathrm{cm}^{-3}$), are listed. For the 
point in time $t_\mathrm{exp}^*$ when the mean shock radius reaches a value of 400\,km, the maximum shock
radius is given, too.

It is common to all simulations presented here that the development of the 
explosion is strongly influenced by the specific density structure of each 
pre-supernova model. All heavier models between 19\,M$_\odot$ and 28\,$\mathrm{M}_\odot$ 
show a pronounced density jump at the interface between the silicon and 
oxygen-enriched silicon (Si/Si-O) shell that is located at radii between 2,000\,km 
and 3,000\,km. While the position of this interface is nearly the same for all 
heavier models, it varies noticeably for the less massive progenitor models and 
in some cases a steep decline in the density profile at the interface cannot be 
observed (see Fig.\,\ref{denplot}). The effects of these different pre-collapse 
structures on the post-bounce evolution  as apparent in our simulations will be 
discussed in depth in the following.   

\subsection{Model Set I}\label{SetI}

\subsubsection{General Properties}\label{general_properties}

\begin{figure}
\includegraphics[width=\columnwidth]{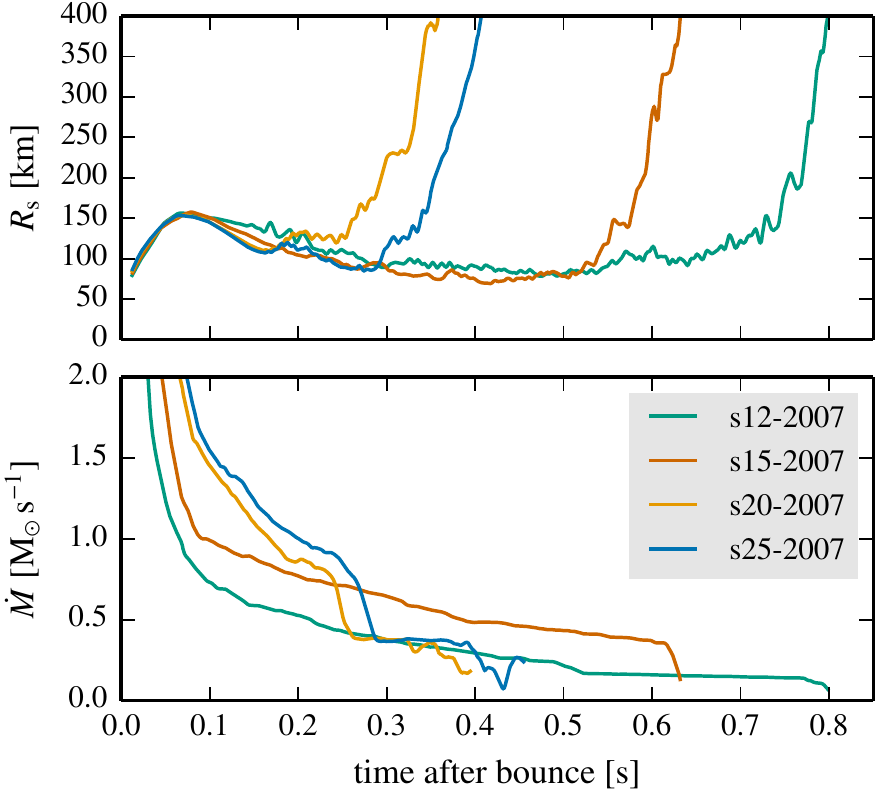}
\caption{Time evolution of average shock radius (upper panel) and mass-accretion 
rate (lower panel) for the simulations of Model Set I. Both quantities are 
averaged over all angular directions; the mass-accretion rate is evaluated at a 
radius of 400\,km. The curves are smoothed by running averages of 
5\,ms.\label{gen_prop}}
\end{figure}

The trajectories of the average shock radii are depicted in Fig.\,\ref{gen_prop} (upper 
panel). All four models explode, but the post-bounce evolution differs. In the case 
of the two more massive progenitors of 20\,M$_\odot$ and 25\,M$_\odot$, the shock retreats 
until it encounters the Si/Si-O composition shell interface. This 
point in time is connected to a steep decrease of the mass-accretion rate 
(evaluated at a radius of 400\,km, see lower panel of Fig. \ref{gen_prop}) given by
\begin{equation}
\dot{M}(r)=4\pi r^2 \rho(r) |v(r)|.
\label{eq:accr}
\end{equation}
Shortly afterwards, the shock starts to expand and the runaway conditions for an 
explosion are reached. This is different in the case of the two progenitors with 
lower masses of 12\,M$_\odot$ and 15\,M$_\odot$. Due to a much weaker density contrast at the 
Si/Si-O  interface, the mass-accretion rate does not show a steep decline. It 
decreases more gradually and the two models explode at relatively late times. The difference 
in the explosion behavior between the two more massive and the two less massive progenitor models can be attributed to the 
competition of mass-accretion rate and neutrino energy deposition in the context of 
the delayed neutrino-driven explosion mechanism. The revival of the stalled 
shock front requires the neutrino heating to be strong enough to overcome the 
ram pressure of the infalling material \citep[e.g.][]{Burrows1993,Janka1996b,Janka2001a,Murphy2008,Fernandez2012}, 
and the threshold conditions for a successful explosion can be defined by a critical neutrino luminosity that 
depends on the mass-accretion rate of the shock \citep{Burrows1993}. We will further
elaborate on this aspect in Sect.\,\ref{Crit}, where we will discuss and demonstrate
the influence of multidimensional fluid flows in the post-shock layer on the critical
luminosity condition in a more general form introduced by \citet{Mueller2015}.

\begin{figure*}
\includegraphics[width=\textwidth]{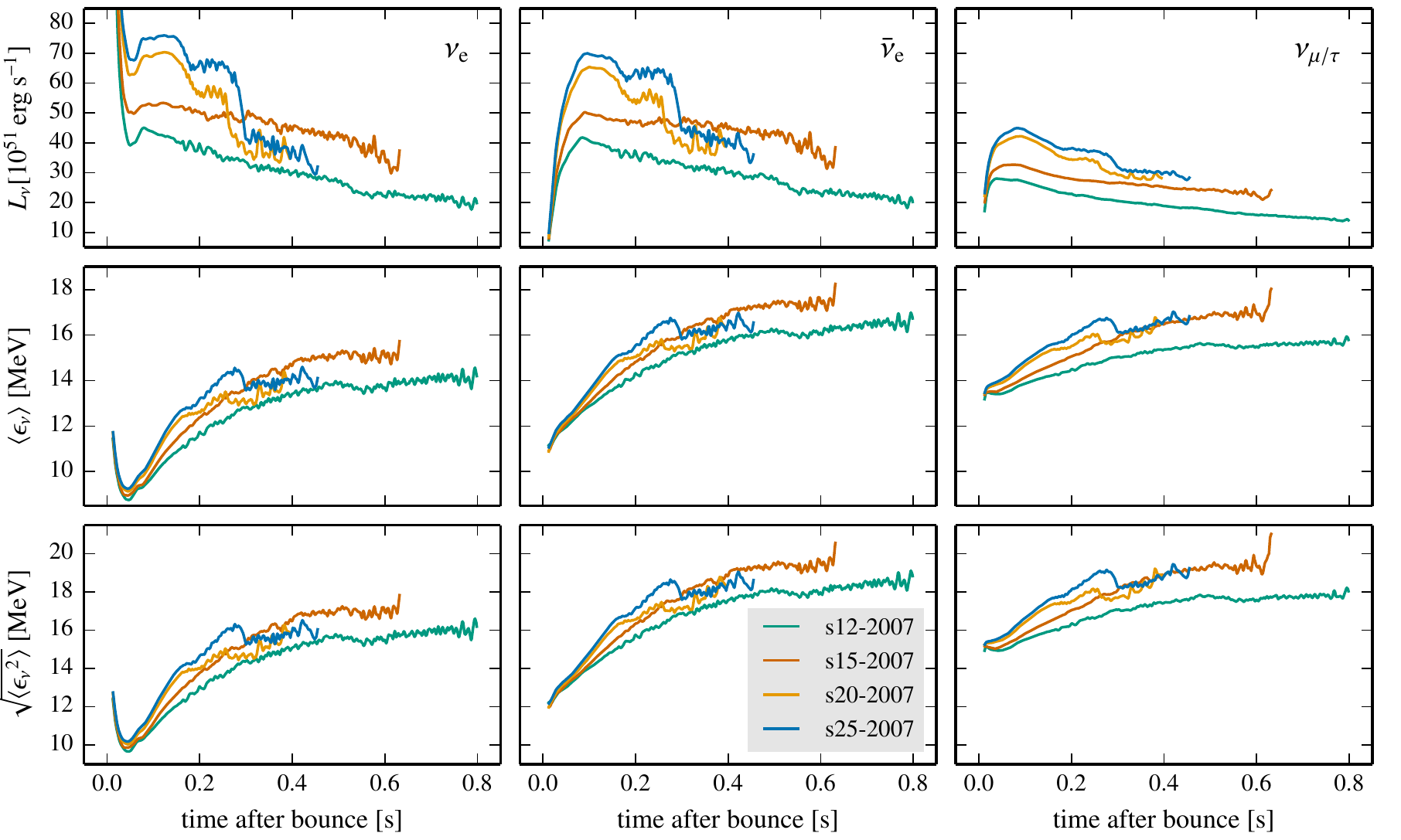}
\caption{Time evolution of neutrino luminosities (first row), neutrino mean 
energies (second row), and neutrino rms energies (third row) for the simulations 
of Model Set I (lab-frame quantities, evaluated at 400\,km and given for an observer
at infinity). From left to right, the angular averages are shown 
for $\nu_\mathrm{e}$, $\bar{\nu}_\mathrm{e}$, and $\nu_{\mu/\tau}$. The curves are smoothed by running averages of 
5\,ms.\label{nu_prop}}
\end{figure*}

In Fig.\,\ref{nu_prop}, the angle-averaged luminosities as well as the angle-averaged mean and rms energies of the 
different neutrino species are shown. These quantities are evaluated at 400\,km and given
for an observer in the lab frame at infinity.
$\langle \epsilon_\nu \rangle$ and $\langle \epsilon_\nu^2 \rangle$ are defined as the first
and second moments of the dimensionless neutrino phase space distribution function $f(\epsilon,\mu)$,
\begin{equation}
\langle \epsilon_\nu^n \rangle = \frac{\int_0^\infty \mathrm{d}\epsilon_\nu\,\epsilon_\nu^n\,\epsilon_\nu^2\, \int_{-1}^1 \mathrm{d}\mu\,f(\epsilon_\nu,\mu)}{\int_0^\infty \mathrm{d}\epsilon_\nu\,\epsilon_\nu^2\, \int_{-1}^1 \mathrm{d}\mu\,f(\epsilon_\nu,\mu)},
\label{eq:nu_ene}
\end{equation}
where $\mu$ is the cosine of the angle between the neutrino momentum and the radial direction
and $\epsilon_\nu$ the neutrino energy. Note that in the multidimensional case the additional
directional averaging involves the integration of the numerator and denominator terms
over all angular directions/bins of the computational grid.

Due to a higher mass-accretion rate and therefore a faster growth of the mass of 
the proto-neutron star, the two more massive models show higher neutrino
luminosities and a faster growth of the radiated mean neutrino energies at early times
of the post-bounce evolution. For this reason, neutrinos deposit more energy in the gain layer and 
provide stronger heating in the region behind the stalled shock. The arrival of the Si/Si-O composition 
shell interface at the shock is reflected by a drop of the neutrino luminosities 
and mean energies, which is further enhanced by the onset of shock expansion (see Fig.\,\ref{nu_prop}). 
At this time, the ram pressure of the 
infalling material is significantly reduced, but a lot of energy is still stored in 
the gain layer behind the shock due to the heating by the previously high accretion 
luminosities. This combination of high neutrino luminosities and mean energies 
but reduced ram pressure is very supportive for the revival of the shock 
\citep[for a detailed discussion, see][]{Ertl2016}.
In the two less massive progenitors, the Si/Si-O interface is relatively weak, and 
during the first 300\,ms after bounce the neutrino luminosities and mean energies are lower. 
Therefore, it takes a longer time until the mass-accretion rate has decreased to 
such a low value that the ram pressure can be overcome by the neutrino heating.

The need for a favorable interplay between neutrino luminosity and mass-accretion 
rate with respect to the onset of a successful explosion is further supported by 
the results of our additional simulations of 14 pre-supernova models 
(Model Set II, see Sect.\,\ref{SetII}) and has been explored in a large set of 1D models
by \citet{Ertl2016}. In the following subsections, we will focus on 
the four simulations of Model Set I and investigate in detail the conditions that 
lead to the initiation of the explosion.

\subsubsection{Conditions in the Gain Layer}\label{gain_layer}

\begin{figure}
\includegraphics[width=\columnwidth]{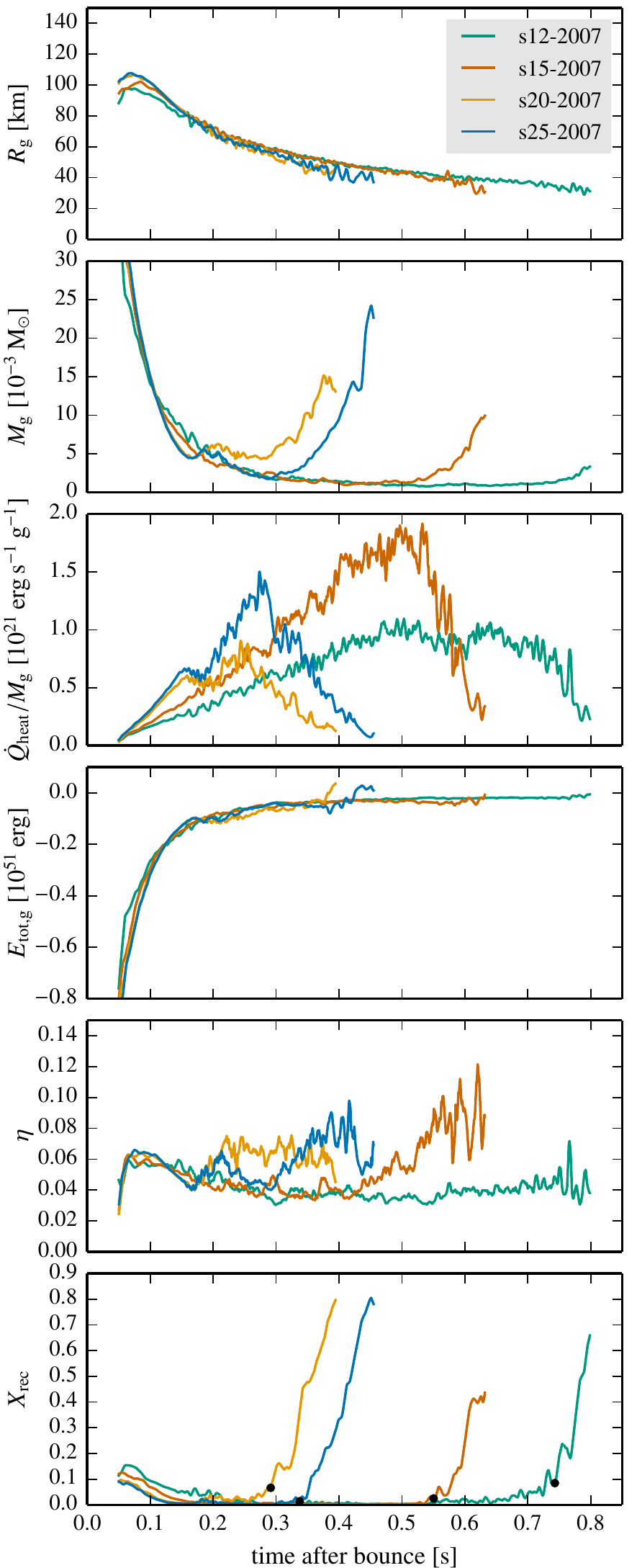}
\figcaption{Time evolution of gain radius, mass, neutrino heating rate per unit mass,
total energy, neutrino heating efficiency, and the mass fraction of
matter recombined to $\alpha$-particles and heavier nuclei in the gain layer (from top to bottom) for the four 
simulations of Model Set I. The black dots in the bottom panel mark the fraction of recombined matter 
at the time when
the ratio $\tau_\mathrm{adv}/\tau_\mathrm{heat}$ reaches unity. Quantities that are not well defined shortly
after bounce are only shown for $t\geq 0.05\,\mathrm{s}$. The curves are smoothed by running averages of 
5\,ms.\label{gain_prop}}
\end{figure}

\begin{figure}
\includegraphics[width=\columnwidth]{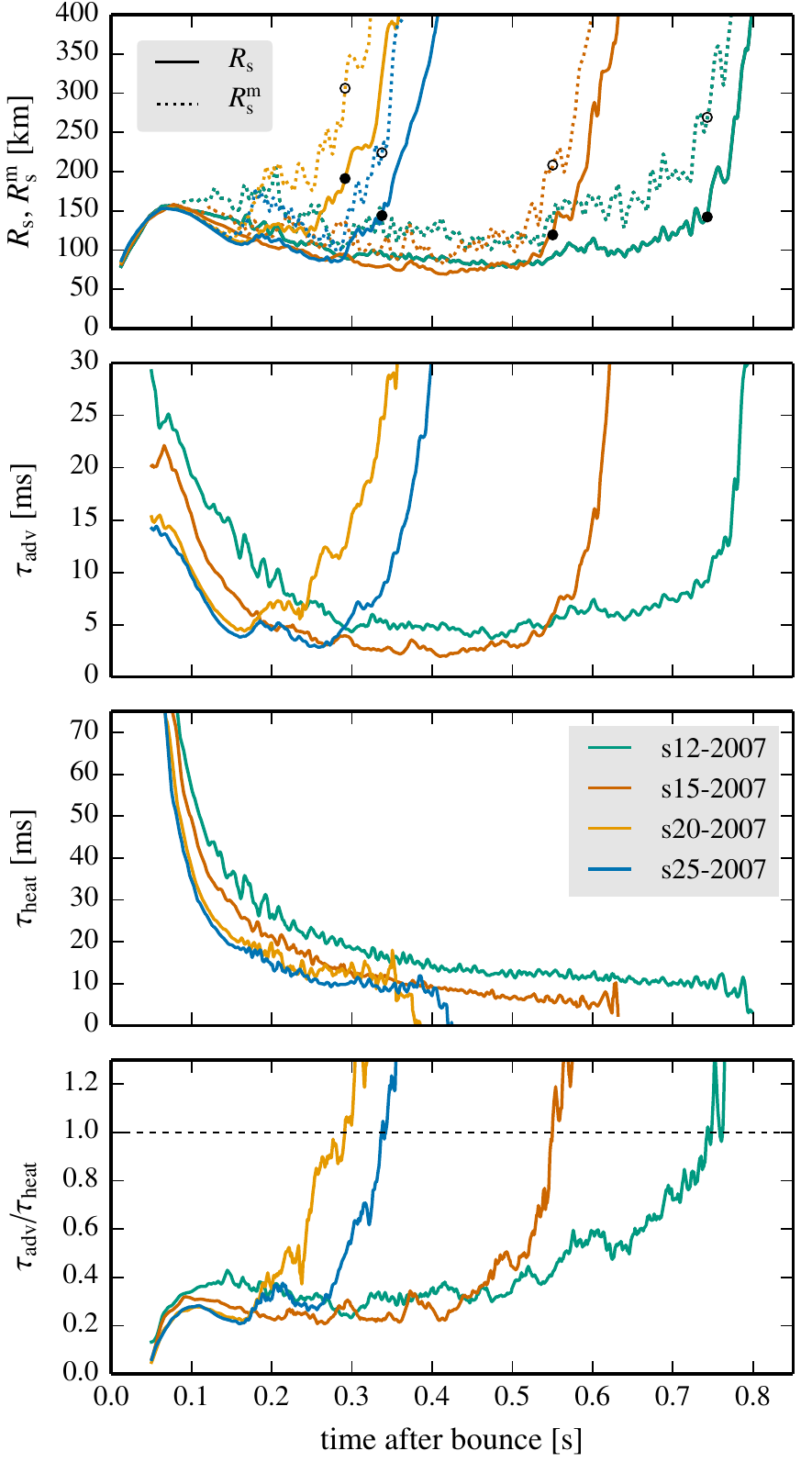}
\caption{Time evolution of average and maximum shock radius,
advection and heating time scale as well as the 
ratio of the two time scales for the four simulations of Model Set I. The definitions of these 
quantities are given in the text. The black dots (open circles) in the top panel mark the value of the averaged (maximum) shock radius 
at the time when
the ratio $\tau_\mathrm{adv}/\tau_\mathrm{heat}$ reaches unity. Quantities that are not well defined shortly
after bounce are only shown for $t\geq 0.05\,\mathrm{s}$. The curves are smoothed by running averages 
of 5\,ms.\label{timescales}}
\end{figure}

The residence time of matter in the gain layer determines the exposure of this 
material to neutrino heating. If the advection time scale 
($\tau_\mathrm{adv}$), defined as the time the accreted gas stays in the gain 
layer, is longer than the heating time scale ($\tau_\mathrm{heat}$), which is given by 
the time neutrino heating needs to deposit an energy equivalent to the binding energy of the 
gas, the conditions in the gain layer become advantageous 
for an explosion. For shock expansion to finally create a runaway 
situation, a sufficiently long period of 
$\tau_\mathrm{adv}/\tau_\mathrm{heat}\gtrsim1$ is necessary 
\citep[e.g.][]{Janka2001b,Thompson2005,Buras2006b,Fernandez2012}. In order to account for the 
influence of non-radial instabilities on the hydrodynamical flow in our 2D 
simulations, we follow \citet{Janka2012a} and \citet{Mueller2012c} and use the 
dwell time of matter in the gain region as a measure for the average advection time scale,
assuming quasi steady-state conditions
\citep[cf.][]{Buras2006b,Marek2009a}:
\begin{equation}
\tau_\mathrm{adv}\coloneqq\tau_\mathrm{dwell}\approx\frac{M_\mathrm{g}}{\dot{M}}.
\label{eq:dwell}
\end{equation}
Here, $\dot{M}$ is the mass-accretion rate through the shock, and 
$M_\mathrm{g}$ is defined by the mass enclosed in the gain layer between the direction-dependent 
(i.e., dependent on the latitudinal angle) gain 
radius $R_\mathrm{g}(\theta)$ and shock radius $R_\mathrm{s}(\theta)$:
\begin{equation}
M_\mathrm{g}=\int \limits_{R_\mathrm{g}(\theta)<r<R_\mathrm{s}(\theta)} 
\rho\,\mathrm{d}V.
\end{equation}
Our definition of the dwell time (Eq.\,\ref{eq:dwell}) is only a 
rough approximation of the advection time scale of matter falling inward through the gain layer,
because this expression also includes material rising with positive
velocities. For exactly this reason, however, Eq.\,(\ref{eq:dwell}) is a good measure of the residence time
of matter in the neutrino-heated region, because the time period of gas being exposed to neutrino heating is increased
by non-radial as well as outward mass motions, which are responsible for a growth of the mass in the gain layer. 
Naturally, after the onset of the explosion, expanding matter begins to dominate in the gain layer, for which
reason Eq.\,(\ref{eq:dwell}) does not yield a good representation of the ``advection time scale'' any longer.

The heating time scale is defined by the ratio of the total energy of the 
material in the gain layer and the volume-integrated neutrino heating rate in 
this region, 
\begin{equation}
\tau_\mathrm{heat}=\frac{\left|E_\mathrm{tot,g}\right|}{\dot{Q}_\mathrm{heat}}.
\label{eq:theat}
\end{equation}
The total energy in the gain layer is given by the integral over the sum of 
specific kinetic energy, $v^2/2$, specific internal energy, $\epsilon$, and specific gravitational binding energy,
\begin{equation}
E_\mathrm{tot,g}=\int \limits_{R_\mathrm{g}(\theta)<r<R_\mathrm{s}(\theta)} \rho \left[ 
\left(\frac{v^2}{2} + \epsilon \right) + \Phi \right] \mathrm{d}V,
\label{eq:ene}
\end{equation}
with $\Phi$ being the gravitational potential.
The neutrino heating rate is the integral of the neutrino energy deposition rate 
per volume $q_\mathrm{e}$ over the gain layer
\begin{equation}
\dot{Q}_\mathrm{heat}=\int 
\limits_{R_\mathrm{g}(\theta)<r<R_\mathrm{s}(\theta)}q_\mathrm{e}\,\mathrm{d}V.
\end{equation}

In Figs.\,\ref{gain_prop} and \ref{timescales}, different diagnostic quantities evaluated for the gain 
region are presented during the post-bounce evolution. While the heating 
time scale continuously decreases with time, the advection time scale shows a 
rapid increase at the time of the arrival of the Si/Si-O interface in the case of the 
two more massive models (see Fig.\,\ref{timescales}). This increase is caused by the sudden 
decline of the mass-accretion rate (cf.\,Fig.\,\ref{gen_prop}, lower panel). The longer residence time of 
matter in the gain region thus enables more efficient neutrino heating 
(Fig.\,\ref{gain_prop}, fifth panel from top) providing 
the power to drive the shock outwards. 

The advection time scale of the two less massive models shows a continuous 
decrease connected to the diminishing amount of mass contained in the gain layer 
(see Fig.\,\ref{gain_prop}, second panel from top) until it stabilizes on a level 
around 5\,ms. Nevertheless, these two models 
still explode at relatively late times after bounce. This can be attributed to 
the increasing heating efficiency (see Fig.\,\ref{gain_prop}, lower panel) defined by the ratio of the total 
energy deposition rate to the sum of the radiated electron neutrino and electron 
antineutrino luminosities (which dominate the heating rate through $\nu_\mathrm{e}$ and 
$\bar{\nu}_\mathrm{e}$ absorption on free nucleons):
\begin{equation}
\eta=\frac{\dot{Q}_\mathrm{heat}}{L_{\nu_\mathrm{e}}+L_{\bar{\nu}_\mathrm{e}}},
\end{equation}
where we measure the luminosities at a radius of 400\,km.
Following \citet{Janka2001a,Janka2012a}, the neutrino energy deposition in the gain 
layer scales with $L_{\nu_\mathrm{e}}$, $L_{\bar{\nu}_\mathrm{e}}$, $\langle E_{\nu_\mathrm{e}}^2\rangle$, 
$\langle E_{\bar{\nu}_\mathrm{e}}^2\rangle$, and $M_\mathrm{g}$ as
\begin{equation}
\dot{Q}_\mathrm{heat}\propto\frac{L_{\nu_\mathrm{e}}\langle 
E_{\nu_\mathrm{e}}^2\rangle+L_{\bar{\nu}_\mathrm{e}}\langle 
E_{\bar{\nu}_\mathrm{e}}^2\rangle}{R^2_\mathrm{g}}M_\mathrm{g}.
\label{eq:qheat}
\end{equation}
Note that $\langle E_{\nu_\mathrm{e}}^2\rangle \coloneqq \langle \epsilon_{\nu_\mathrm{e}}^3\rangle / \langle \epsilon_{\nu_\mathrm{e}}\rangle$ 
and $\langle E_{\bar{\nu}_\mathrm{e}}^2\rangle \coloneqq \langle \epsilon_{\bar{\nu}_\mathrm{e}}^3\rangle / \langle \epsilon_{\bar{\nu}_\mathrm{e}}\rangle$
are defined from the energy distribution of neutrinos
in energy space, not from the number distribution as $\langle \epsilon_{\nu_\mathrm{e}}^2\rangle$ 
and $\langle \epsilon_{\bar{\nu}_\mathrm{e}}^2\rangle$.
Since the mass in the gain layer (see Fig.\,\ref{gain_prop}, second panel from 
top) is growing at later times and the neutrino rms energies (see Fig.\,\ref{nu_prop}, third row)
are continuously increasing, too, the slow decline of the accretion luminosity (see 
Fig.\,\ref{nu_prop}, first row) can be overcompensated and the heating 
efficiency rises noticeably already before the onset of the explosion 
\citep[cf.][]{Marek2009a,Mueller2012c}. This effect can be observed in the two 
less massive models: At late times, $\nu_\mathrm{e}$ and $\bar{\nu}_\mathrm{e}$ deposit a larger fraction of 
their energy in the gain layer, the post-shock flow is heated more efficiently and 
finally an explosion is triggered.

\begin{figure}
\includegraphics[width=\columnwidth]{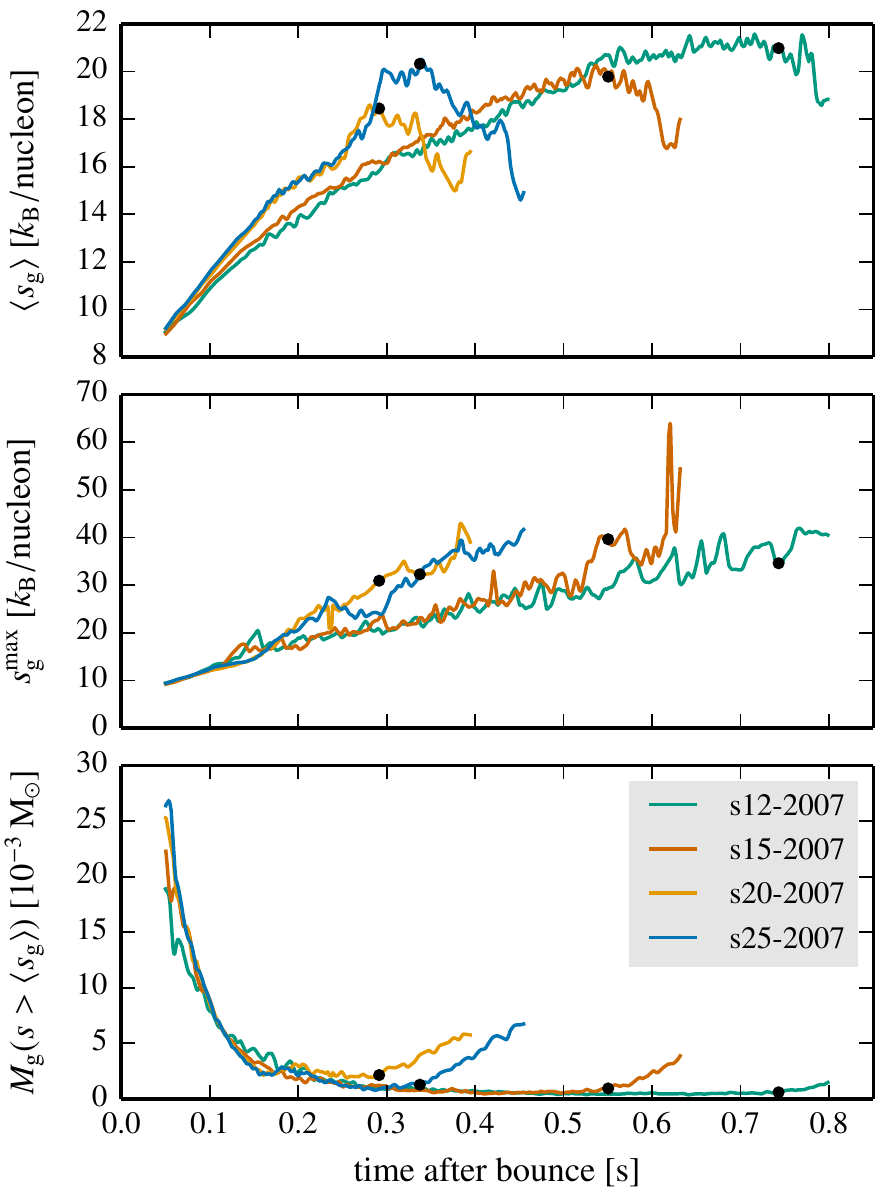}
\caption{Time evolution of average and maximum entropy in the gain layer as well as the mass in the gain layer
with entropies above the average value (from top to bottom). The black dots mark the point in time when
the ratio $\tau_\mathrm{adv}/\tau_\mathrm{heat}$ reaches unity. The quantities are shown for 
$t\geq 0.05\,\mathrm{s}$ and the curves are smoothed by running averages 
of 5\,ms. 
\label{stoplot}}
\end{figure}

Around the onset of the explosion, the advection time scale rises steeply in all 
four models. Higher pressure and stronger ``turbulent'' flows 
in the gain layer lead to an expansion of gas outward 
from deeper layers of the gain region. The expansion of the shock creates a 
positive feedback loop by further increasing the advection time scale. Once the 
critical condition of $\tau_\mathrm{adv}/\tau_\mathrm{heat}\gtrsim1$ is reached, 
a runaway situation with continuous shock expansion is created \citep[e.g.][]{Buras2006a,Murphy2008,Fernandez2012}. 
The evolution of the total energy in the gain layer can be inferred from Fig.\,\ref{gain_prop} 
(fourth panel from top). When the time-scale ratio 
$\tau_\mathrm{adv}/\tau_\mathrm{heat}$ reaches unity, the total energy is still slightly
negative \citep[compare Figs.\,\ref{gain_prop} and \ref{timescales}; cf.\ also][]{Janka2001a,Fernandez2012}. At the beginning 
of the shock expansion, just a small fraction of the material in the gain 
layer is rising while most parts of the matter behind the shock are still nearly at 
rest (see also Sect.\,\ref{instabilities}). Only when the whole gain layer 
starts to expand, does the total energy tend towards positive values indicating that 
the post-shock material gets unbound in the gravitational field created by the enclosed mass. 
In Fig.\,\ref{stoplot}, the average entropy in the gain layer, defined by
\begin{equation}
\langle s_\mathrm{g} \rangle = \frac{1}{M_\mathrm{g}}\int\limits_{R_\mathrm{g}(\theta)<r<R_\mathrm{s}(\theta)} s\,\rho\, \mathrm{d}V,
\end{equation}
as well as the maximum entropy and the mass in the gain layer with entropies above the 
average value are given.
In all four models, the entropy increases towards explosion. On the way to explosion also the mass
in the gain layer with entropies above the average value is growing, which is compatible with
previous findings that the masses and volumes with entropies above certain threshold values grow
\citep{Nordhaus2010a,Hanke2012,Fernandez2014}.
Although models exploding at later times after bounce show a tendency towards higher entropies, no generic 
value that signals the successful runaway can be found.
Once the explosion has started, a great amount of lower-entropy gas from below the gain radius enters
the gain layer and leads to a drop of the average entropy by $2-4\,k_\mathrm{B}$ per nucleon.

Overall, the time-scale criterion seems to be a viable concept for interpreting
the explosion behavior of all four models. We will provide further evidence for that
in Sect.\,\ref{Crit}. The concept is post-dictive in the sense
that it is based on an analysis of the model conditions (in contrast to a two-parameter criterion found
by \citet{Ertl2016}, which is based on the properties of the pre-collapse star). There is ongoing controversy in the 
literature whether better explosion indicators exist that describe the approach to runaway shock
expansion in a physically more founded way \citep[e.g.][]{Pejcha2012,Murphy2015,Gabay2015}.
We do not want to take a position in the debate here because we focus on
multidimensional results while the cited literature discusses the behavior of the shock-stagnation
problem in 1D, where special pathologies like large-scale radial shock pulsations can occur, which
do not have a direct counterpart in 2D and 3D. \citet{Janka2012a} and \citet{Mueller2015} have
shown that the critical condition of the time-scale ratio can formally be connected to the 
critical luminosity condition $L_\mathrm{crit}(\dot{M})$ introduced by \citet{Burrows1993}; 
see also, e.g., \citet{Yamasaki2005}, \citet{Murphy2008}, \citet{Nordhaus2010a}, \citet{Hanke2012}, and \citet{Fernandez2012}, which 
can be generalized to include the effects of non-radial fluid flows in terms of a contribution
by turbulent pressure \citep[cf.][]{Mueller2015}. In Sect.\,\ref{Crit}, we follow the approach of
\citet{Mueller2015} and demonstrate that a generalized critical condition can be formulated
that applies to the whole set of 18 models as a general criterion for the onset of the explosion.

In multidimensional simulations, the development towards a runaway 
situation is closely connected to the evolution of hydrodynamic instabilities.
The growth conditions of these instabilities are the topic of the 
next subsection, where their properties are further discussed in dependence on the 
different progenitor models.

\subsubsection{Growth of Instabilities}\label{instabilities}

\begin{figure}
\includegraphics[width=\columnwidth]{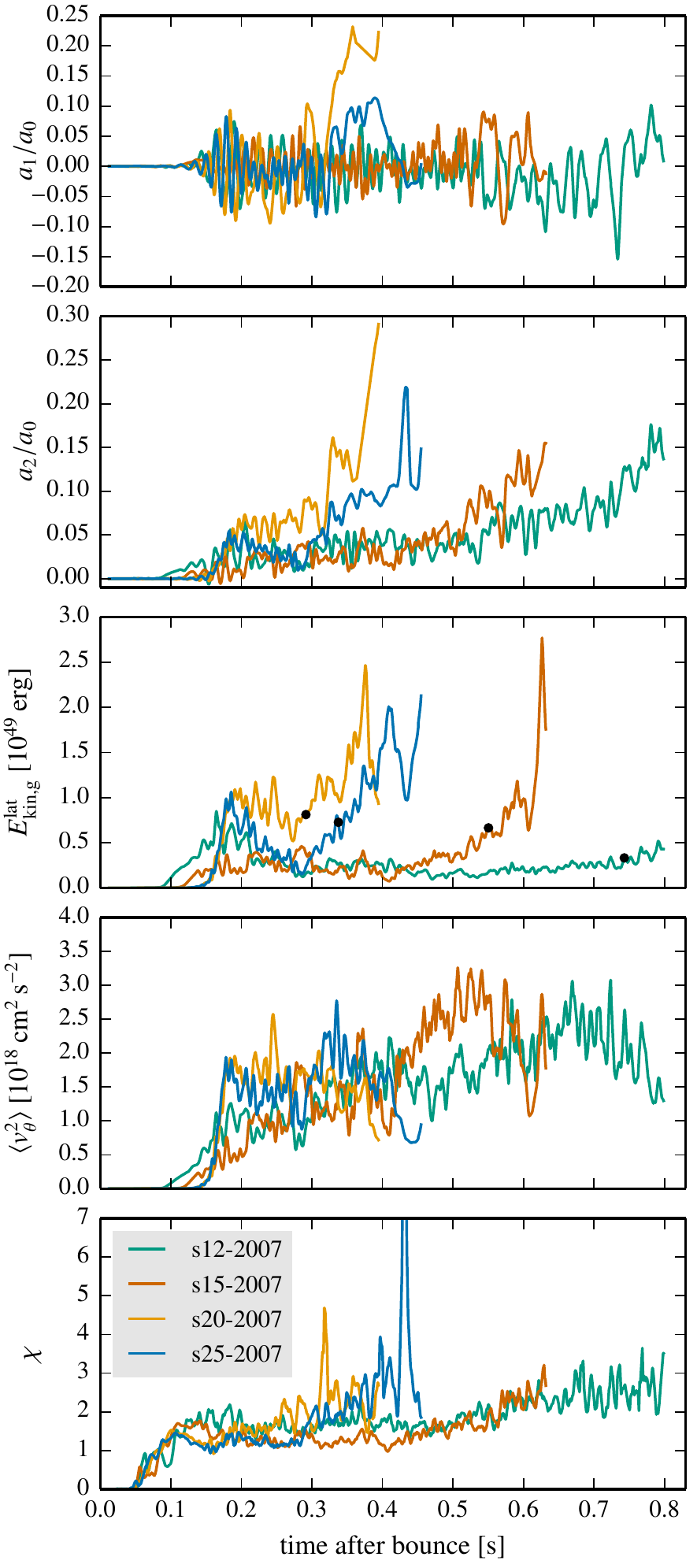}
\caption{Time evolution of the coefficients $a_1$ (dipole mode, first panel) and 
$a_2$ (quadrupole mode, second panel) for an expansion of the shock surface into 
Legendre polynomials. The coefficients are normalized to the amplitude of the 
$l=0$ mode (i.e.\,the average shock radius). Furthermore, the kinetic energy of lateral 
mass motions (third panel) and the velocity dispersion in the gain layer (fourth 
panel) are given. The black dots in the third panel mark the kinetic energy level at the time when
the ratio $\tau_\mathrm{adv}/\tau_\mathrm{heat}$ reaches unity.
In the bottom panel, the growth parameter $\chi$ is depicted 
for the four simulations of Model Set I. All curves are smoothed by running 
averages of 5\,ms.\label{inst}}
\end{figure}

    \begin{figure*}
	    \subfigure{
            \includegraphics[width=0.455\textwidth]{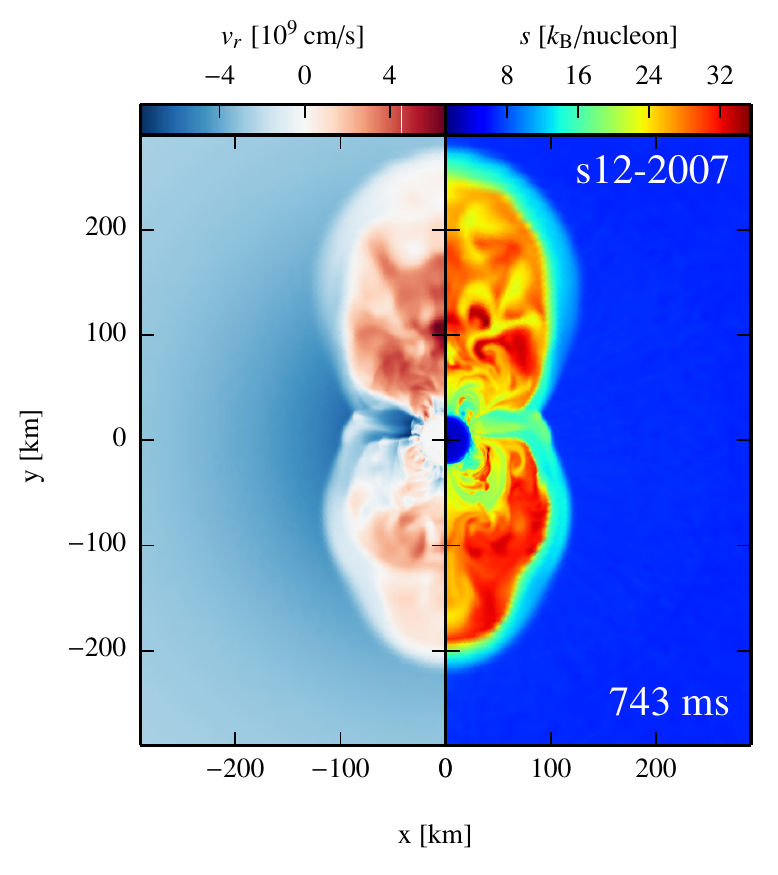}
}
        \quad
	    \subfigure{  
            \includegraphics[width=0.455\textwidth]{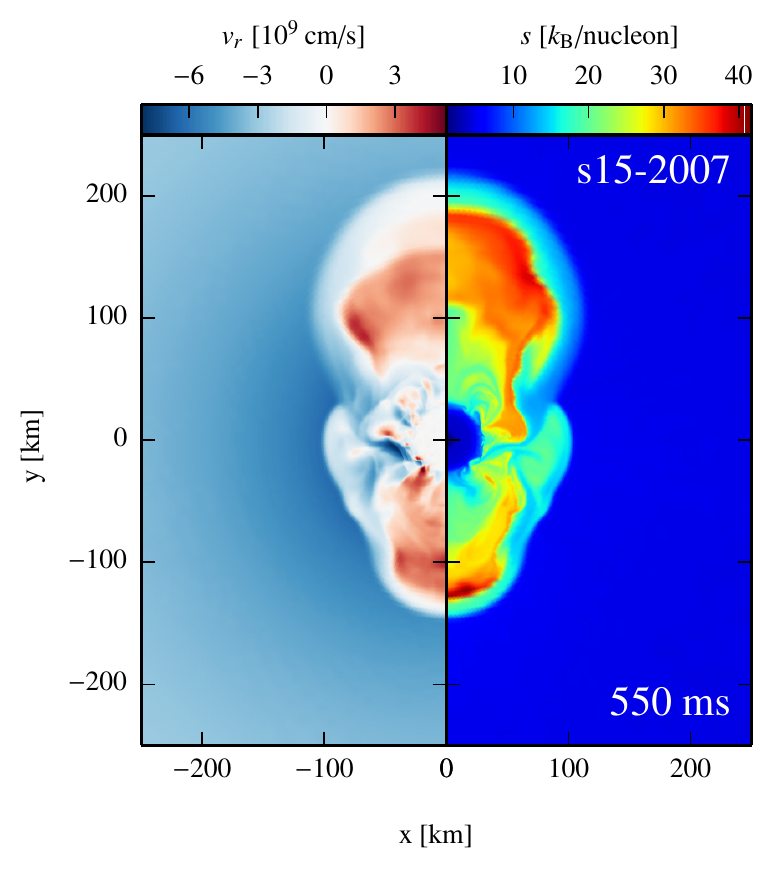}
}
        \vskip -0.5cm
	    \subfigure{
            \includegraphics[width=0.455\textwidth]{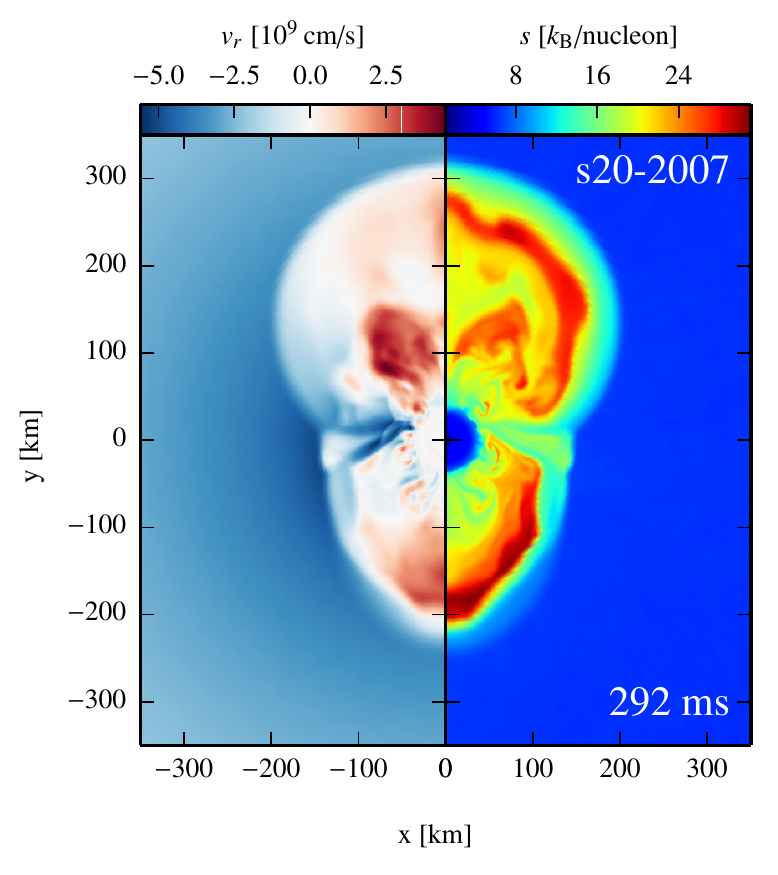}
}
        \quad
	    \subfigure{
            \includegraphics[width=0.455\textwidth]{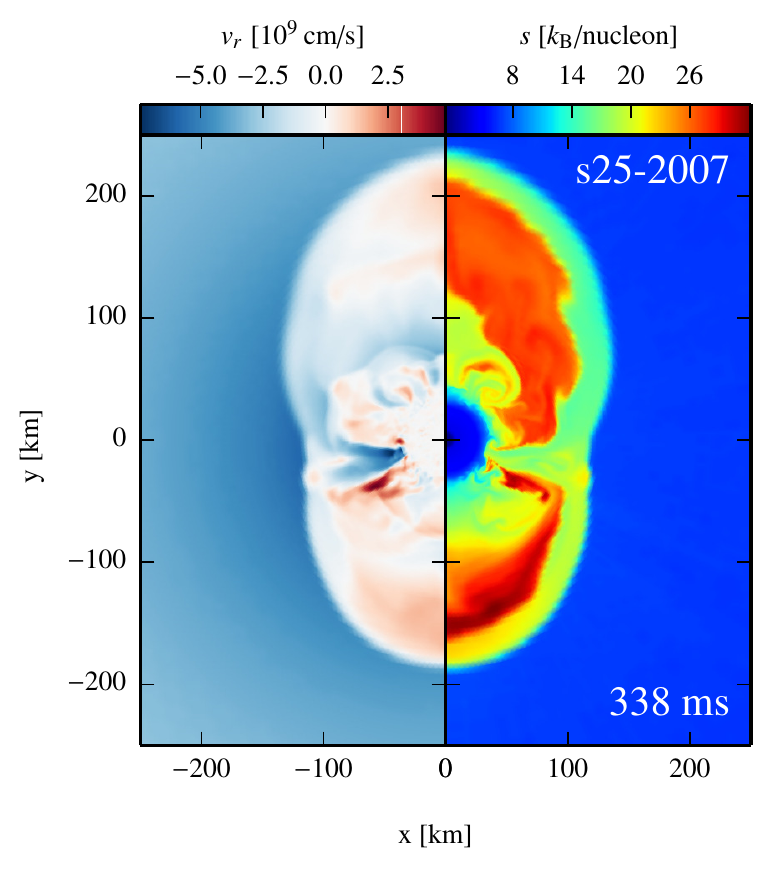}
}
	\vspace{-0.25cm}
        \caption{Snapshots of radial velocity (left halfs of the panels) and entropy per 
baryon (right halfs of the panels) for the four simulations of Model Set I at the time of 
explosion (defined by the time when the ratio 
$\tau_\mathrm{adv}/\tau_\mathrm{heat}$ reaches unity).} 
        \label{2Dplots}
    \end{figure*}

Non-radial mass motions are crucial for an increase of the dwell time of matter 
in the gain layer, enhanced neutrino heating, turbulent pressure, and the 
subsequent expansion of the shock radius \citep{Murphy2013,Mueller2015}. Both 
convection and the standing accretion-shock instability \citep[SASI,][]{Blondin2003} 
can provide sufficient support to the neutrino-heating mechanism to finally revive the previously 
stalled shock front. Typically, the high mass-accretion rates of the investigated 
models (see Fig.\,\ref{gen_prop}, lower panel) and the decreasing neutron star radii 
(see Fig.\,\ref{exp}, second panel) lead to very small shock radii stabilizing at $\sim 
80\,\mathrm{km}$, well following the proportionality
\begin{equation}
R_\mathrm{s}\propto \frac{\left(L_\nu\langle E_{\nu_\mathrm{e}}^2\rangle\right)^{4/9}R_\mathrm{NS}^{16/9}}{\dot{M}^{2/3}M_\mathrm{NS}^{1/3}}
\label{eq:rs}
\end{equation}
\citep{Janka2012a} as soon as quasi-steady state accretion conditions in the post-shock
layer apply\footnote{A quasi-stationary state is earliest reached after the shock has arrived at its maximum radius,
because the initial shock expansion is driven by the high mass-accretion rate, which leads to a non-stationary accumulation of an accretion 
mantle around the neutron star core. This is demonstrated in Appendix\,\ref{evo_shock}, where we show the mass-accretion rates and shock trajectories
for our Model Set I.} and as long as multidimensional effects do not play a crucial role
\citep[for a generalization to multi-dimensions, see Eq.\,(\ref{eq:r_shock_turb}) below and][]{Mueller2015}.
Due to the scaling relation \citep[cf.][]{Scheck2008a}
\begin{equation}
\tau_\mathrm{adv}\propto R_\mathrm{s}^{3/2},
\label{eq:adv}
\end{equation}
the advection time scale shrinks 
accordingly. The linear growth rate $\omega_\mathrm{SASI}$ of the advective-acoustic 
cycle amplifying the SASI growth is given by \citep{Foglizzo2006}
\begin{equation}
\omega_\mathrm{SASI}=\frac{\mathrm{ln}|\mathcal{Q}|}{\tau_\mathrm{cyc}},
\end{equation}
where $\mathcal{Q}$ is the efficiency and $\tau_\mathrm{cyc}$ is the duration of 
the cycle. As argued by \citet{Scheck2008a} and \citet{Mueller2012}, $\tau_\mathrm{cyc}$ is short for 
small shock radii and thus short advection time scales. Hence, our models with 
rather short advection time scales should provide favorable conditions for 
efficient SASI growth.
In order to quantify this expected behavior by a detailed analysis, we decompose 
the angle-dependent shock surface $R_\mathrm{s}(\theta)$ into Legendre 
polynomials $P_l(\cos\theta)$. The expansion coefficients are defined by 
\citep{Burrows2012,Ott2013}
\begin{equation}
a_l=\frac{1}{2}\int 
\limits_{0}^{\pi}R_\mathrm{s}(\theta)\,P_l(\cos\theta)\,\mathrm{d}
(\cos\theta).
\end{equation}

For the four models of Model Set I, the time evolution of the coefficient $a_1$ 
(dipole mode) is shown in Fig.\,\ref{inst} (upper panel). At $\sim 120\,\mathrm{ms}$ 
after bounce (average shock radii between 120\,km and 150\,km, see Fig.\ref{gen_prop}, first panel), 
shock sloshing motions begin to grow in the well-known oscillatory way. At this 
time, the lateral kinetic energy in the gain region increases (see Fig.\,\ref{inst}, third panel from top) 
and the post-shock flow becomes aspherical. All models exhibit strong 
quasi-periodic shock oscillations with oscillation periods of 15\,ms to 20\,ms. At 
$\sim 220\,\mathrm{ms}$ after bounce, the Si/Si-O composition shell interface 
reaches the shock in the case of the two more massive models, and the advection 
time scale and hence the SASI oscillation period increase \citep{Foglizzo2007,Scheck2008a,Guilet2012}. 
During the shock expansion phase, the two models still show large shock 
oscillations, but these oscillations are less regular than before. In the two 
less massive models, shock oscillations with short periods can be maintained up 
to several hundred milliseconds after bounce. Only after the shock expansion 
sets in, can larger SASI amplitudes with non-periodic behavior and large shock excursions
with unipolar or bipolar asymmetry be observed.

The time evolution of the quadrupole mode represented by the coefficient $a_2$ 
is depicted in Fig.\,\ref{inst} (second panel from top). Shortly before shock 
expansion sets in, all four models develop a growing prolate quadrupolar 
deformation of the shock surface. In the two more massive models that develop 
explosions at earlier times, the $a_2$ coefficient almost continuously increases 
directly from the onset of SASI activity at $\sim 120\,\mathrm{ms}$. In the case of 
the less massive models exploding at later times, the increase of the quadrupole 
mode starts at $\sim 450\,\mathrm{ms}$. The development of a strong quadrupole 
mode in all four models is a serious hint that the artificial symmetry axis 
introduced in 2D simulations may play a supportive role for the runaway 
expansion of the shock \citep[e.g.][]{Takiwaki2012,Hanke2012,Couch2013}. 
The quadrupole mode periodically pushes the post-shock 
layer further and further out towards the polar direction along the symmetry axis, while 
inflow occurs along funnels near the equator. The big polar, buoyant bubbles are
fed by material from equatorial downflows, which channel accreted matter to the
gain radius, where it can be efficiently heated by neutrinos.

The large-amplitude bipolar oscillations with increasing amplitudes 
push the shock front step by step 
outwards to larger radii. Due to the effective increase of the dwell time of matter 
that is channelled into the polar lobes, more accreted 
material can be heated by the neutrinos for longer times (see bottom panel of 
Fig.\,\ref{gain_prop} for the growing neutrino heating efficiency). The continuously 
ongoing SASI oscillations successively drive the shock front outwards, which in 
turn further increases (cf. Eq.\,\ref{eq:dwell}) the advection time scale of matter 
in the gain layer. This positive feedback loop finally induces a successful explosion 
\citep{Marek2009a,Mueller2012c}.

Additionally, the supportive role of the SASI for shock revival is mirrored in 
supersonic lateral velocities (sound speed $c_\mathrm{s} \sim 10^9\,\mathrm{cm\,s}^{-1}$) 
in the post-shock flow caused by repeated 
phases of large-amplitude shock expansion and contraction. The kinetic energy of 
these non-radial mass motions shows quasi-periodic variations with spiky maxima 
(see Fig.\,\ref{inst}, third panel from top). As pointed out by \citet{Hanke2012}, this is typical of 
the presence of low-order SASI modes. Similar to the results of their parametric 
study for models at the explosion threshold, the successful explosions presented 
here are triggered and accompanied by large-scale mass flows, which are indicated by growing 
fluctuations of the angular kinetic energy that are characteristic for strong 
SASI activity. 

While the lateral kinetic energy also depends on the mass contained in the gain 
region, the velocity dispersion $\langle v_\theta^2\rangle$ provides a direct 
measure for the typical velocities of convective and SASI motions and for the 
turbulent pressure associated with them \citep{Mueller2015}. Consequently, 
the continuous growth of this quantity for all models indicates an increase of 
convective and SASI activity with time. This is especially supportive for the 
development of an explosion at several hundred milliseconds after bounce in the case 
of the two less massive models. Following \citet{Mueller2015}, the lateral 
kinetic energy satisfies the relation
\begin{equation}
\frac{E_\mathrm{kin,g}^\mathrm{lat}}{M_\mathrm{g}}\propto \left[\left(\langle R_\mathrm{s}\rangle-\langle R_\mathrm{g}\rangle\right) 
\frac{\dot{Q}_\mathrm{heat}}{M_\mathrm{g}}\right]^{2/3}.
\label{eq:ekin}
\end{equation}
Since the neutrino heating rate per unit of mass, $\dot{Q}_\mathrm{heat}/M_\mathrm{g}$, 
scales with $L_{\nu_\mathrm{e}} \langle E_{\nu_\mathrm{e}}^2\rangle + L_{\bar{\nu}_\mathrm{e}} \langle E_{\bar{\nu}_\mathrm{e}}^2\rangle$ 
(cf.\,Eq.\,\ref{eq:qheat} and see Fig.\,\ref{gain_prop}, third panel from top), 
the continuous increase of the mean neutrino energies is also 
responsible for the growth of the velocity dispersion and fosters the 
large-scale aspherical mass motions which finally induce the onset of 
explosion. 

While the conditions of the hydrodynamic post-shock flow are favorable for the 
efficient development of the SASI in our simulations, convection is generally 
suppressed. Similar to the results of \citet{Scheck2008a} and 
\citet{Marek2009a}, the neutrino energy deposition in the gain layer of our 
models is too weak to generate a steep negative entropy gradient. The latter is a 
prerequisite for the development of convection. Furthermore, the applied 
EoS of \citet{Lattimer1991} with a nuclear incompressibility of 
220\,MeV generates rather compact neutron stars \citep{Steiner2010,Hebeler2010}. 
Thus, the forming neutron stars contract rapidly from a maximum radius of 
$\sim 75\,\mathrm{km}$ to $\sim 40\,\mathrm{km}$ after 200\,ms post bounce and
$\sim 25\,\mathrm{km}$ after $\sim 800\,\mathrm{ms}$ post bounce. Since the shock radius 
directly scales with the neutron star radius (measured by the radial location of
$\rho=10^{11}\,\mathrm{g\,cm}^{-3}$; see Eq.\,\ref{eq:rs}), the contraction 
of the neutron star also enforces the retraction of the shock radius. That is 
why the matter in the post-shock region is rapidly advected towards the gain 
radius and the growth of convective motions is suppressed \citep[cf.][]{Foglizzo2006}.

In order to quantify the importance of convection, we determine the growth 
parameter $\chi$ introduced by \citet{Foglizzo2006} for our four explosion 
models (see Fig.\,\ref{inst}, lower panel). This parameter can be considered as a measure of the 
ratio of the advection time scale of the flow through the gain layer and the 
growth time scale of convection. It is defined in terms of the 
Brunt-V\"ais\"al\"a frequency $\langle\omega_\mathrm{BV}\rangle$  
\citep[calculated from angle-averaged quantities, for a discussion see][]{Fernandez2014} and the spherically averaged 
advection velocity $\langle v_r\rangle$ by
\begin{equation}
\chi=\int \limits_{\langle R_\mathrm{g}\rangle}^{\langle 
R_\mathrm{s}\rangle}\frac{\mathrm{Im}\,\langle\omega_\mathrm{BV}\rangle}{|\langle v_r 
\rangle|}\mathrm{d}r,
\end{equation}
where the integration runs from the averaged gain radius to the averaged shock 
radius. Only regions with $\omega_\mathrm{BV}^2<0$ (indicating local instability) 
contribute to the integral. Since perturbations are advected out of the gain 
layer with the accretion flow in a finite time, a sufficient amplification of 
initial perturbations within this time interval is needed for the successful 
development of convective motions. According to the analysis of 
\citet{Foglizzo2006} in the linear regime of small initial perturbations, a 
threshold condition of $\chi\gtrsim3$ is necessary for convective instability in 
the gain region. This condition is compatible with several numerical studies 
in 2D \citep[e.g.][]{Buras2006b,Scheck2008a,Fernandez2009,Fernandez2014}.

While SASI activity starts at around $\sim 100\,\mathrm{ms}$ after bounce 
when aspherical mass motions begin to develop (see Fig.\,\ref{inst}, two upper panels), the growth 
parameter for convective instability still remains subcritical (see Fig.\,\ref{inst}, lower panel). 
Due to the low neutrino heating rates and the small shock radii and 
correspondingly short advection time scales at 
these times, convection is damped in all four models. This absence of 
convection may also be supportive for the early development of the SASI 
\citep[cf.][]{Mueller2012}. In the case of the two more massive models, the 
threshold condition of $\chi>3$ is reached after the Si/Si-O composition shell 
interface has arrived at the shock. Because of the abruptly reduced mass 
accretion rate, the shock expands to larger radii and the advection time scale 
rises. This leads to increased values of $\chi$. 

The two less massive models retain a subcritical value of $\chi$ for a long 
time. After $\sim 500\,\mathrm{ms}$, the conditions for convection become more 
and more favorable. Convective activity is fully established when the time-scale 
ratio $\tau_\mathrm{adv}/\tau_\mathrm{heat}$ exceeds unity and shock expansion 
sets in. The reason for the gradual development of convection in these two 
models is two-fold. On the one hand, the large-amplitude SASI sloshing motions 
of the stalled shock front are associated with fast lateral flows in the post-shock region 
(see Fig.\,\ref{inst}, third and fourth panel from above) and induce the formation of layers with very steep unstable entropy 
gradients \citep[see also][]{Scheck2008a,Marek2009a}. This supports the emergence of secondary 
convective activity \citep{Buras2006b,Scheck2008a}. On the other hand, the 
increasing values of the $\chi$ parameter directly mirror the enhanced neutrino energy 
deposition per unit of mass (see Fig.\,\ref{gain_prop}, third panel from top) at late times.

While our simulations show a similar behavior as the ``SASI-dominated'' model 
s27.0 presented by \citet{Mueller2012}, a clear disentanglement of SASI and 
convective effects with respect to the post-shock dynamics emerging around shock 
revival is difficult. In the case of strongly aspherical flows due to the SASI, with
perturbations far away from the linear regime, the 
criterion $\chi>3$ may no longer be a reliable measure for the development of 
convective instability. To illustrate the hydrodynamic properties of the 
post-shock flow around shock revival, color-coded snapshots of entropy and radial 
velocity are presented in Fig.\,\ref{2Dplots} for all models at the time when the time-scale 
ratio $\tau_\mathrm{adv}/\tau_\mathrm{heat}$ reaches unity. All models show a 
prolate deformation of the shock surface caused by large-amplitude bipolar SASI 
oscillations. 
In addition to small buoyant bubbles growing in the wake of the SASI 
sloshing motions, large-scale high-entropy bubbles triggered by the SASI shock 
expansion phases are visible.
Due to the assumption of axisymmetry, large plumes preferentially grow
along the direction of the artificial symmetry axis \citep[see also][]{Takiwaki2012,Hanke2012,Couch2013}.

\begin{figure}
\includegraphics[width=\columnwidth]{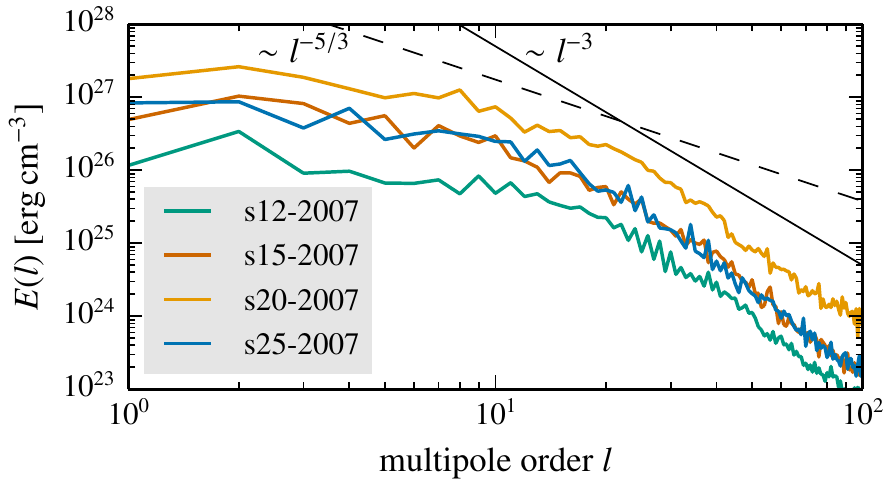}
\caption{Turbulent energy spectra $E(l)$ as functions of the multipole order $l$. 
The spectra are based on a decomposition of the azimuthal velocity $v_\theta$ into
spherical harmonics 25\,ms before the time-scale ratio $\tau_\mathrm{adv}/\tau_\mathrm{heat}$ 
exceeds unity, averaged over a time interval of 5\,ms and a radius interval from 50\,km to 80\,km. The thin dashed
and solid black lines indicate reference spectral slopes of $-5/3$ and $-3$.}\label{turb}
\end{figure}

According to \citet{Fernandez2014}, SASI-dominated explosion models are 
characterized by the interplay of shock sloshing motions and the formation of 
large-scale, high-entropy structures. The authors conclude that a SASI-driven 
explosion develops if these bubbles are able to survive during several SASI 
oscillation periods. The dominance of large-scale bubbles seeded by SASI 
sloshing motions compared to small-scale bubbles driven by convection as 
indicated by the snapshots in Fig.\,\ref{2Dplots} clearly suggests that the post-shock flow 
dynamics in our simulations are governed by the SASI while convective 
instabilities play a more secondary role. This interpretation is further supported by
an analysis of the energy spectrum $E(l)$ which considers the decomposition of the
azimuthal velocity $v_\theta$ at a given radius (weighted by the square root 
of the density) into spherical harmonics 
$P_{l}(\cos\theta)$ as the 2D analogon of the definition provided by \citet{Hanke2012}:
\begin{equation}
E(l)= \frac{1}{2}\left| \sqrt{\left(2l+1\right)\pi}\int \limits_0^\pi P_{l}\left(\cos\theta\right)\sqrt{\rho}\,v_\theta\left(r,\theta\right)\mathrm{d}(\cos\theta)\right|^2.
\end{equation}
The results of this analysis are shown in Fig.\,\ref{turb}. In order to obtain smoother
spectra, $E(l)$ is averaged over 30\,km in radius and over 5\,ms in time. Similar to 
the SASI-dominated models discussed by \citet{Fernandez2014}, the angular spectrum of 
all four models shows a peak at $l=2$. The strong presence of convection is visible
from the enhanced power in the $l=5-10$ domain fully compatible with the spectral features
observed in the convection dominated models by \citet{Fernandez2014}. This confirms the 
fact that the $\chi$ parameter tends towards the critical value of 3 or even begins to 
exceed this value when the time-scale ratio approaches unity.
The slope of $\sim -3$ at large $l$ is indicative 
for a direct vorticity cascade being characteristic of the spectral properties of
turbulence in axisymmetry \citep{Kraichnan1967}. 

\subsubsection{Diagnostic Explosion Energies and Neutron Star Properties}

\begin{figure}
\includegraphics[width=\columnwidth]{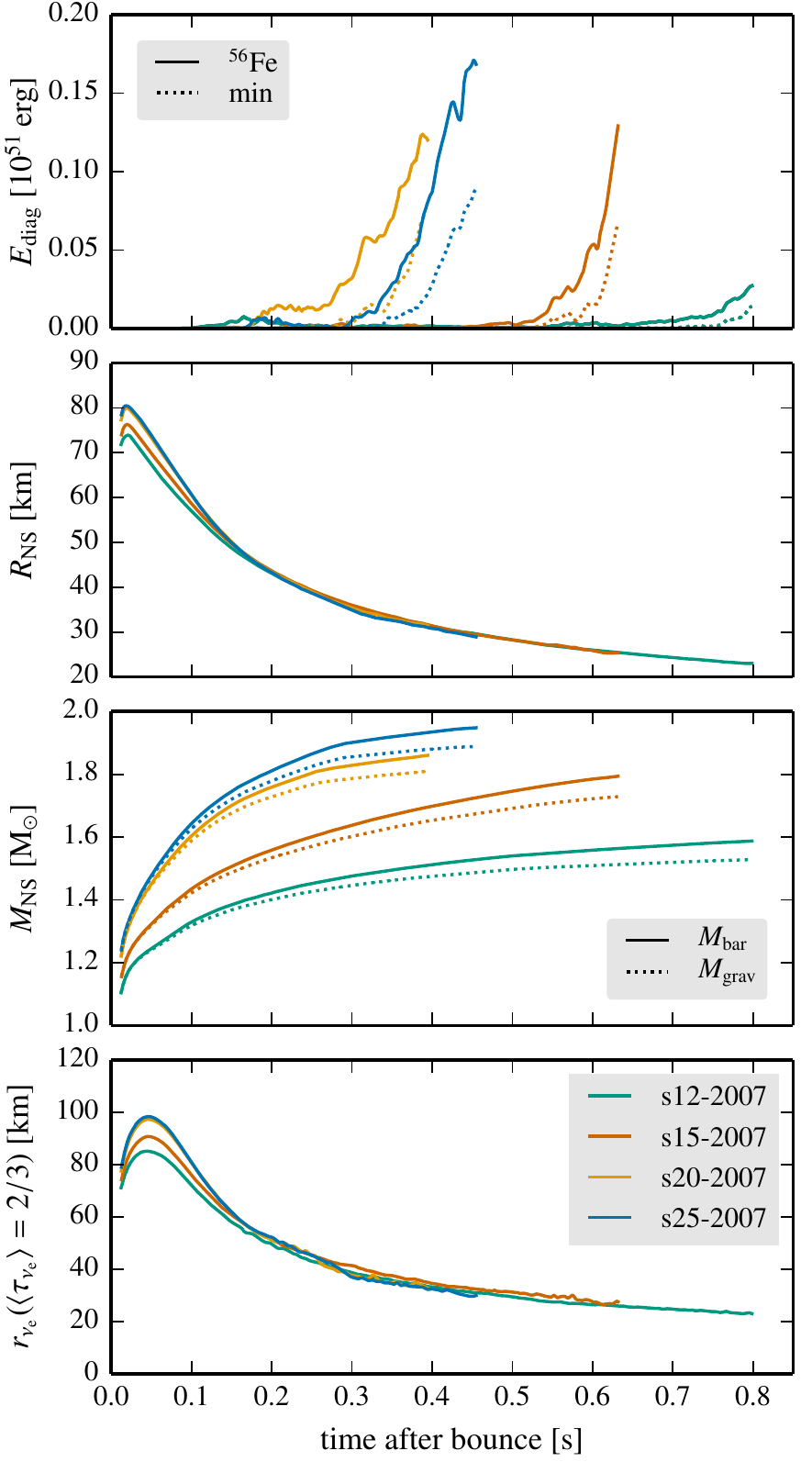}
\caption{Time evolution of diagnostic energy (lower limits indicated by dotted 
lines, upper limits by solid lines, see text), neutron star radius and mass (the baryonic mass is
denoted by solid lines, the gravitational mass by dotted lines), and 
the radius of the spectrally averaged $\nu_\mathrm{e}$ sphere at an optical depth of $\langle\tau_{\nu_\mathrm{e}}\rangle=2/3$ 
(from top to bottom). Neutron star radius and mass are defined by the density 
surface at $10^{11}\,\mathrm{g}\,\mathrm{cm}^{-3}$. All quantities are 
angle-averaged and the curves are smoothed by running averages of 
5\,ms.\label{exp}}
\end{figure}

The diagnostic energies depicted in Fig.\,\ref{exp} (dotted lines in 
upper panel) are calculated by integrating over the gain layer for regions
where the total specific energy defined as $e_\mathrm{tot}=v^2/2+\epsilon+\Phi$ is positive 
\citep[see also previous studies 
by][]{Buras2006a,Marek2009a,Suwa2010,Mueller2012c,Bruenn2013}:
\begin{equation}
E_\mathrm{diag}=\int \limits_{R_\mathrm{g}(\theta)<r<R_\mathrm{s}(\theta)} 
e_{\mathrm{tot,>0}}\,\rho\,\mathrm{d}V.
\end{equation}
In addition to this lower limit, we also show diagnostic explosion energies 
obtained by the assumption that all nucleons finally recombine to iron-group 
nuclei, which accounts for the maximum release of nuclear binding energy and can 
be considered as upper limit (see Fig.\,\ref{exp}, solid lines in upper panel).
Typically, first fluid elements behind the shock become formally unbound ($e_\mathrm{tot}>0$) at the onset of 
the explosion when the time-scale ratio exceeds unity. After the shock has 
expanded beyond $\sim 200\,\mathrm{km}$, the temperature behind the shock 
decreases sufficiently to allow for the recombination of nucleons to 
$\alpha$-particles (see bottom panel of Fig.\,\ref{gain_prop} for the fraction $X_\mathrm{rec}$ 
of recombined matter in the gain layer). Consequently, the explosion energy starts to rise with a 
steep gradient. At the time our simulations had to be stopped because of the 
extremely high computational demands of the neutrino transport, maximum 
diagnostic energies of up to $\sim0.17\times 10^{51}\,\mathrm{erg}$ were reached 
and were still increasing steeply. 

However, at this stage of the simulations a reliable determination of the final 
explosion energies is not possible. 
In order to follow the energy budget of unbound matter and the continuous 
recombination processes behind the expanding shock front, the simulations would 
have to be carried on further for several hundred milliseconds 
\citep[cf.][]{Scheck2006a,Scheck2008a}. This is presently beyond reach due to
extremely small transport time steps. Because of ongoing accretion and mass ejection
we expect that the explosion energies can rise considerably even after the onset of the 
explosion \citep[cf.][]{Marek2009a,Mueller2012c,Mueller2015a}.

The time evolution of the baryonic and gravitational\footnote{The gravitational neutron star mass is directly derived from the
effective general relativistic potential described in \citet{Marek2006a}, which is identical
to subtracting the time-integrated total neutrino luminosity from the baryonic mass.} neutron star masses and radii defined by the density 
surface at $10^{11}\,\mathrm{g}\,\mathrm{cm}^{-3}$ as well as the radius of the 
spectrally averaged electron neutrino sphere at an (effective) optical 
depth of $\langle\tau_{\nu_\mathrm{e}}\rangle=2/3$ is shown in 
the three lower panels of Fig.\,\ref{exp}. For computing the optical depth
for neutrino equilibration we used the effective opacity 
\begin{equation}
\kappa_\mathrm{eff} = \sqrt{\kappa_\mathrm{tot}\,\kappa_\mathrm{abs}},
\end{equation}
where $\kappa_\mathrm{abs}$ is the opacity for neutrino absorption processes and 
$\kappa_\mathrm{tot} = \kappa_\mathrm{abs} + \kappa_\mathrm{scatt}$ is the total
opacity for absorption and scattering. The preliminary value of the neutron star 
mass is determined by the amount of matter that can be accreted from the collapsing
star and settles to densities above $10^{11}\,\mathrm{g\,cm}^{-3}$ until the 
end of our simulations. After the strong decrease of the mass-accretion rate caused 
by the arrival of the Si/Si-O interface in the two more massive models, the increase of 
the neutron star masses begins to flatten. The higher growth rate of the neutron 
star mass in model s15-2007 compared to model s12-2007 directly 
reflects the differences of the mass-accretion rates in these two simulations 
that persist until the explosions set in at late times (compare Fig.\,\ref{eq:accr}).  

\subsection{Model Set II}\label{SetII}

\begin{figure*}
\includegraphics[width=\textwidth]{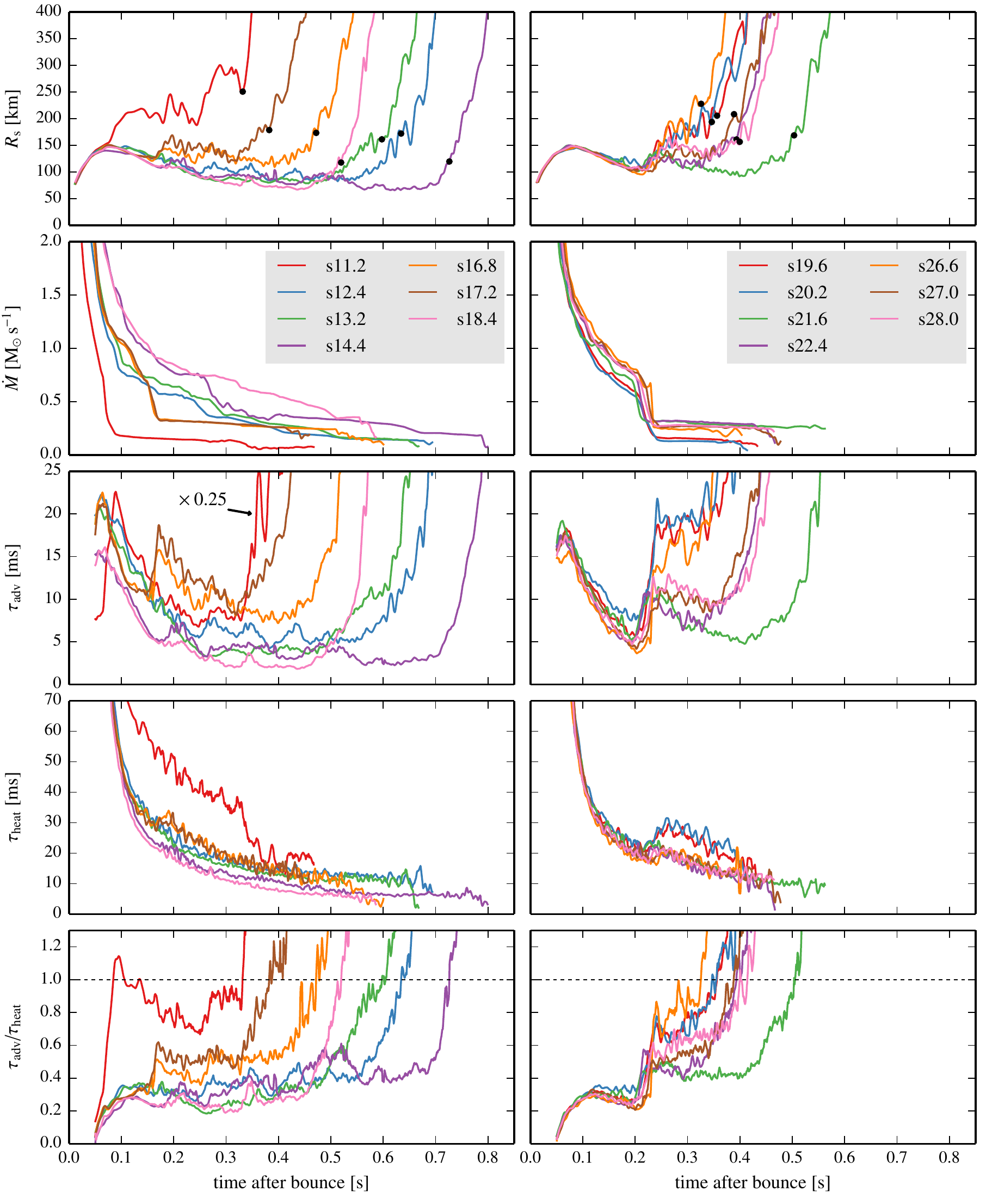}
\caption{Time evolution of different diagnostic quantities for Model Set II. For 
better clarity, the 14 models are subdivided into two parts: the models with 
lower ZAMS masses are displayed in the left column, the models with higher ZAMS 
masses in the right column. From top to bottom, the average shock radius, the 
mass-accretion rate, the advection time scale, the heating time scale, and the
time-scale ratio are depicted. Quantities that are not well defined shortly after bounce are 
only shown for $t\geq0.05\,\mathrm{s}$ post bounce. The black dots in the top panels
mark the point in time when the ratio $\tau_\mathrm{adv}/\tau_\mathrm{heat}$
reaches unity (in the case of model s11.2, the transient spike of $\tau_\mathrm{adv}/\tau_\mathrm{heat}\gtrsim1$
at early times was disregarded). 
Note that the advection time scale of model s11.2 (left column,
third panel from top) is scaled with a factor of 0.25. All quantities are angle-averaged and 
the curves are smoothed by running averages of 5\,ms.
\label{set2_a}}
\end{figure*}

\begin{figure*}
\includegraphics[width=\textwidth]{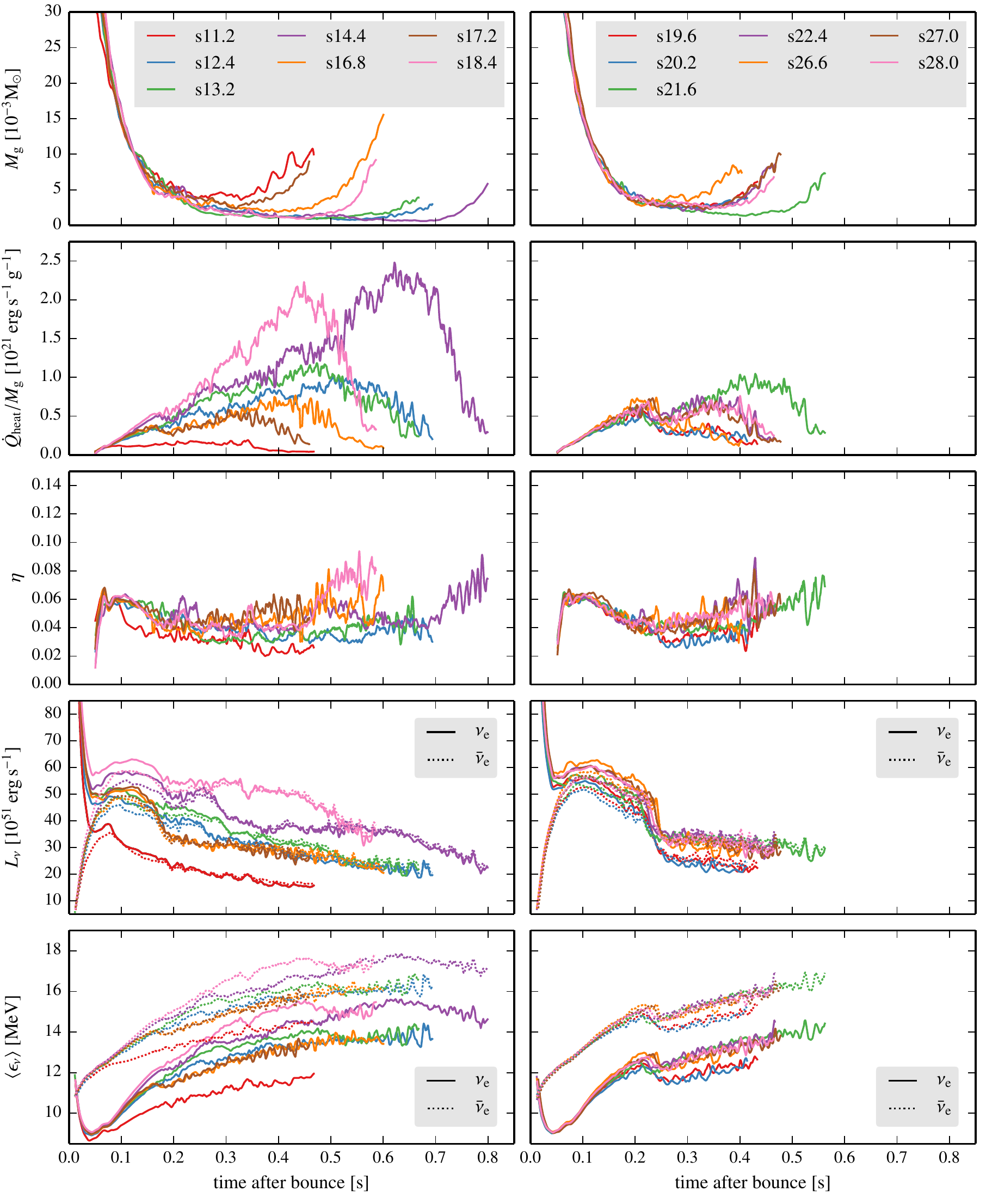}
\caption{Time evolution of different diagnostic quantities for Model Set II. For 
better clarity, the 14 models are subdivided into two parts: the models with 
lower ZAMS masses are displayed in the left column, the models with higher ZAMS 
masses in the right column. From top to bottom, the mass in the gain layer, the 
neutrino heating rate per unit mass, the neutrino heating 
efficiency, and the luminosities and mean energies of electron neutrinos (solid lines)
and electron antineutrinos (dotted lines) are depicted. Quantities that are not well defined shortly after bounce are 
only shown for $t\geq0.05\,\mathrm{s}$ post bounce. All quantities are angle-averaged and 
the curves are smoothed by running averages of 5\,ms.
\label{set2_b}}
\end{figure*}

In the following, the main results of our simulations (Set II) concerning 14 
pre-supernova models of \citet{Woosley2002} are presented in the light of the
preceding discussion of Set I. An overview of the characteristic 
properties of these models is given in Figs.\,\ref{set2_a} and \ref{set2_b}. 

The differences in the position and density gradient of the Si/Si-O interface (see 
Fig.\,\ref{denplot}) are directly mirrored by the temporal evolution of the mean shock 
radii of the models with lower and higher ZAMS masses 
(see Fig.\,\ref{set2_a}, first row). The most outstanding examples are models 
s19.6, s20.2, and s26.6 with a very pronounced jump of the density at the interface. 
After the arrival of this jump at the shock surface, the shock almost 
continuously expands outwards. The time evolution of these models is comparable to
that of models s20-2007 and s25-2007 extensively discussed in Sect.\,\ref{SetI}. 
For model s21.6, the delay between the 
arrival of the interface and the beginning of the shock expansion is largest, because for 
this model the step-like decrease of the mass-accretion rate is less extreme than in 
the other representatives of the subset of more massive models 
(see Fig.\,\ref{set2_a}, second row). The less massive stars that do not 
show a sharp discontinuity at the Si/Si-O interface (especially the 12.4\,M$_\odot$, 13.2\,M$_\odot$, 
14.4\,M$_\odot$, and 18.4\,M$_\odot$ cases) explode only at relatively late times 
when the mass-accretion rates have decreased sufficiently, 
similar to the models s12-2007 and s15-2007 of Model Set I. 

Model s11.2 which has already been intensively studied in previous works 
\citep{Buras2006b,Marek2009a,Mueller2012c,Suwa2013} can be considered as special case. In 
this model, the Si/Si-O composition shell interface arrives already at 
$\sim80\,\mathrm{ms}$ after bounce and at this time, the mass-accretion rate 
decreases to a much lower value ($\sim0.2\,\mathrm{M}_\odot\,\mathrm{s}^{-1}$) than in 
the other less massive models. This is why the shock front can expand to large 
radii at early times. In spite of a transient overshoot of $\tau_\mathrm{adv}/\tau_\mathrm{heat}=1$
at $\sim 100\,\mathrm{ms}$ post bounce, however, the 11.2\,M$_\odot$ model explodes only
when this critical value of the time-scale ratio is exceeded for a long-lasting period later
than $\sim 300\,\mathrm{ms}$ after bounce \citep[see also][]{Marek2009a}.

In general, the trends already discussed in the previous section for the four 
explosion models of \citet{Woosley2007} also hold for the 14 models of 
\citet{Woosley2002}. The major prerequisites for a relatively 
immediate onset of the explosion can be summarized as follows. High mass-accretion rates
and proto-neutron star masses at the time before the Si/Si-O interface
reaches the shock surface cause high neutrino luminosities and mean energies. This leads 
to strong neutrino heating, which still persists when the interface has passed the shock 
front. The more pronounced the density jump at the interface is, the lower the mass-accretion 
rate gets and therefore the ram pressure of the infalling material, producing
very favorable conditions for a successful shock revival.

In cases of models exploding at relatively late times (e.g.\ s12.4, s13.2, s14.4, s18.4, and s21.6), 
a stabilization of $\tau_\mathrm{adv}/\tau_\mathrm{heat}$ at values well below 
unity can be observed (see Fig.\,\ref{set2_a}, bottom row). Nevertheless, these models 
still achieve to explode after a longer accretion phase. The systematically 
increasing neutrino heating rates per unit mass (see Fig.\,\ref{set2_b},
second row from top) result in a continuous growth of the velocity dispersion 
in the gain layer (cf. Eq.\,\ref{eq:ekin}), supporting the 
development of strong hydrodynamical instabilities, which are crucial for the final rise of the 
time-scale ratio above unity (cf. Sect.\,\ref{instabilities}).

On the whole, our self-consistent axisymmetric simulations of Model Set II with 
\textsc{Prometheus-Vertex} fully confirm the strong dependence of the 
explosion characteristics on the specific progenitor structure as already 
concluded in the investigation of Model Set I.  

\section{A Generalized Approach Towards the Critical Neutrino Luminosity Condition}\label{Crit}

\begin{figure*}
\centering
\includegraphics[width=1.25\columnwidth]{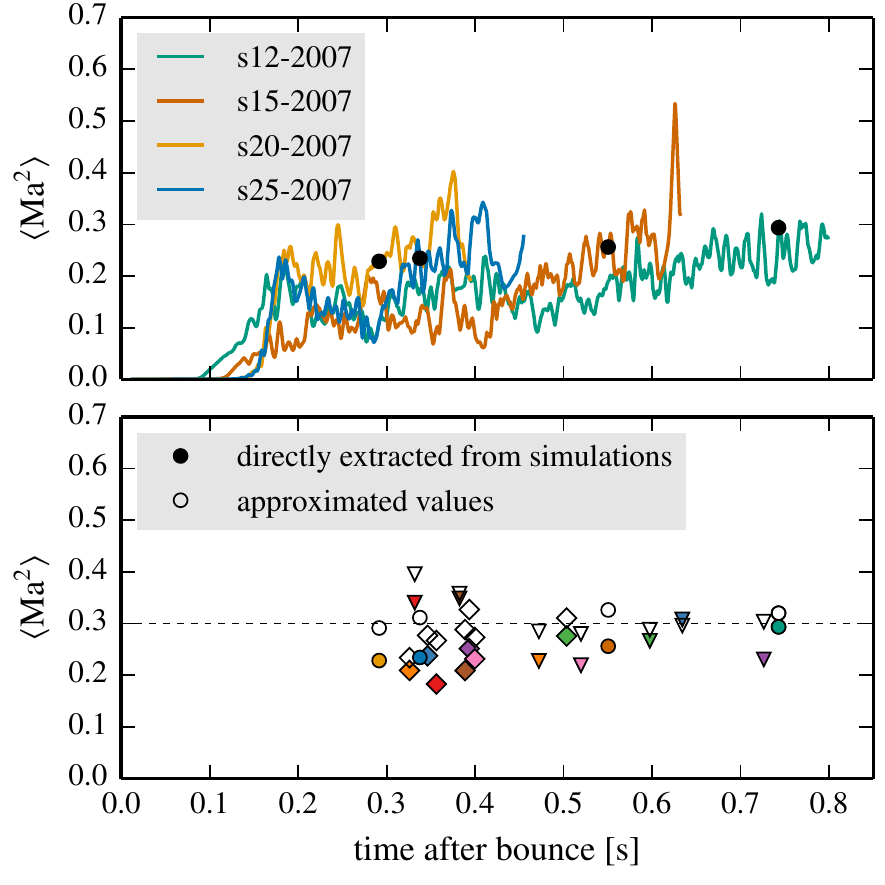}
\caption{In the upper panel, the time evolution of the average squared Mach number in the gain layer
(calculated without approximation) is shown for the simulations of Model Set I. 
The curves in the upper panel are smoothed by running averages of 5\,ms.
In the bottom panel, the average squared Mach number (cf.\,Eq.\,\ref{eq:mach}) 
is given for all 18 models at the time when the ratio $\tau_\mathrm{adv}/\tau_\mathrm{heat}$ 
reaches unity. Filled symbols indicate the values directly evaluated from
the simulations, empty symbols denote the approximation given by Equations (41) and (52) of \citet{Mueller2015}. 
The dashed line indicates the critical limit suggested by \citet{Mueller2015}.
\label{mach}}
\end{figure*}

Although the time-scale criterion appears to be a reliable concept for the
description of the explosion behavior in all 18 axisymmetric simulations 
(in a post-dictive, diagnostic manner), at first glance
no obvious correlations with other characteristic quantities 
can be found that point to generally valid properties at the onset of the explosion. At 
the time the ratio $\tau_\mathrm{adv}/\tau_\mathrm{heat}$ reaches unity, the
models exhibit a diverse range of average and maximum shock radii, 
neutrino luminosities and mean energies, kinetic energies
and fractions of recombined matter in the gain layer, etc. (see, for example, Tab.\,\ref{tab1} 
and Figs.\,\ref{gen_prop} to \ref{stoplot}), and the conditions necessary for shock
revival do not seem to be constrained tightly enough to define a common 
framework for a successful runaway. 

\citet{Mueller2015} suggest that in 2D a squared turbulent Mach number of $\langle \mathrm{Ma}^2\rangle \gtrsim 0.3$
is needed for runaway.
The average squared Mach number of the turbulent lateral motions in the gain region is defined as
\begin{equation}
\langle\mathrm{Ma}^2\rangle = \frac{\langle v_\theta^2 \rangle}{\langle c_\mathrm{s,g}^2\rangle} = \frac{2E_\mathrm{kin,g}^\mathrm{lat}/M_\mathrm{g}}{\langle c_\mathrm{s,g}^2\rangle}.
\label{eq:mach}
\end{equation}
In contrast to \citet{Mueller2015}, we do not employ further approximations for the sound speed 
$c_\mathrm{s,g}$, but extract all quantities directly from the numerical simulations
as mass-weighted averages over the gain layer instead of quantities measured behind the shock:
\begin{equation}
\langle c_\mathrm{s,g}^2 \rangle = \frac{1}{M_\mathrm{g}}\int\limits_{R_\mathrm{g}(\theta)<r<R_\mathrm{s}(\theta)} c_\mathrm{s}^2 \rho\, \mathrm{d}V.
\end{equation}

The consequences of the two different approaches concerning the determination of $c_\mathrm{s,g}$
can be inferred from Fig.\,\ref{mach}, where the time evolution of the average squared Mach number for Model Set I 
is shown (calculated without approximation, see upper panel)
and the average squared Mach numbers of all 18 models are given at the time the ratio $\tau_\mathrm{adv}/\tau_\mathrm{heat}$
reaches unity (calculated with and without approximation, see lower panel). 
The time scales $\tau_\mathrm{adv}$ and $\tau_\mathrm{heat}$ are calculated according 
to Eqs.\,(\ref{eq:dwell}) and (\ref{eq:theat}).
While the approximate calculation of $\langle\mathrm{Ma}^2\rangle$ 
only takes into account post-shock quantities with a number of simplifying assumptions 
\citep[see][]{Mueller2015}, the direct calculation 
considers the (averaged) properties of the whole gain region, because such an analysis offers more numerical
robustness than a calculation directly behind the shock. The latter approach typically results in smaller Mach numbers 
(compare empty and filled circles in the lower panel of Fig.\,\ref{mach}), and the correlation between Mach number and 
the onset of explosion (defined by $\tau_\mathrm{adv}/\tau_\mathrm{heat}=1$) points towards a
`critical squared Mach number' around $\sim0.25$ and thus below the value of 0.3 found by \citet{Mueller2015}.
However, there are considerable temporary fluctuations in which $\langle\mathrm{Ma}^2\rangle$ can
exceed the value of 0.25 for transient times even before shock runaway occurs. Moreover, at the time when 
$\tau_\mathrm{adv}/\tau_\mathrm{heat}\sim1$, the individual values scatter by more than $\sim30\,\%$ around
the mean critical value of all models, for which reason the turbulent Mach number is at most indicative,
but has no hard threshold for shock runaway. This suggests a considerable model-to-model variation of the 
turbulent pressure contribution, being only one of several elements that play a role in triggering the 
explosion.

\begin{figure*}
\centering
\includegraphics[width=1.25\columnwidth]{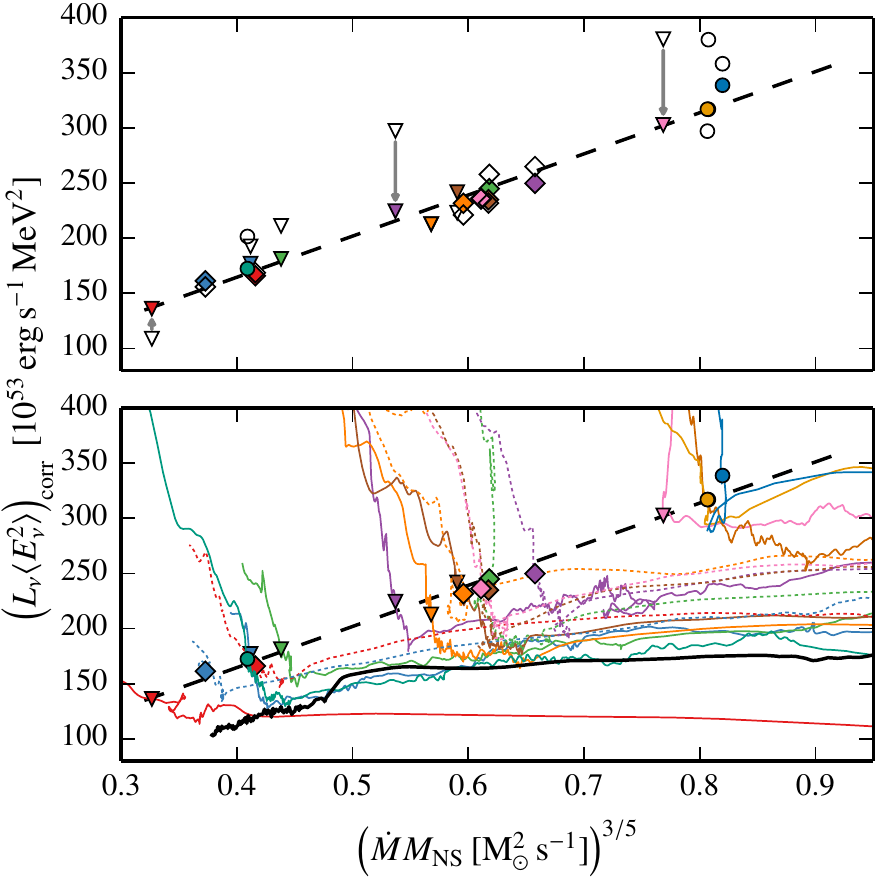}
\caption{Critical luminosity condition for explosion. In both panels, 
the critical relation between $\left(L_\nu\langle E_\nu^2\rangle\right)_\mathrm{corr}$
and $\left(\dot{M}M_\mathrm{NS}\right)^{3/5}$ for the onset of explosion (see Eqs.\,\ref{eq:lnu_crit} and \ref{crit_cond}) 
is depicted as black dashed line obtained from a least-squares fit to the critical points 
of all 18 axisymmetric models.
The symbols indicate these critical points corresponding to the time when the ratio $\tau_\mathrm{adv}/\tau_\mathrm{heat}$ reaches unity. 
Circles denote the models of Set I, and triangles (diamonds) indicate the models of Set II with lower (higher) ZAMS 
masses following the classification and color coding used in previous figures. In the upper panel, a comparison between corrected 
(filled symbols) and uncorrected (empty symbols) values of the critical luminosity is shown. In three exemplary cases, the shifts 
introduced by the correction factor are indicated by gray arrows. In the lower panel, the trajectories of all 18 models in the 
$\left(L_\nu\langle E_\nu^2\rangle\right)_\mathrm{corr}$-$\left(\dot{M}M_\mathrm{NS}\right)^{3/5}$ plane for corrected values 
are additionally given (models of Set II with higher ZAMS masses are 
shown with dotted lines). Furthermore, an axisymmetric (2D)
$15\,\mathrm{M}_\odot$ model of \citet{Heger2005} that does not evolve towards an explosion is depicted with a 
black solid line. All depicted values are smoothed by running averages of 25\,ms.\label{corr}}
\end{figure*}

In principle, it is possible to relate the `critical luminosity' \citep{Burrows1993,Murphy2008,Pejcha2012}
that is required to overcome the ram pressure at a given mass-accretion rate
to the time-scale criterion $\tau_\mathrm{adv}/\tau_\mathrm{heat}\gtrsim1$
\citep[cf.][]{Janka2012a}. But in contrast to studies in spherical symmetry, non-radial
instabilities in multidimensional simulations play a crucial role for the supernova 
explosion mechanism and directly influence the critical luminosity condition.
While theories have been proposed to describe the saturation properties of the 
SASI \citep[e.g.][]{Guilet2010} and of convection \citep[e.g.][]{Murphy2011,Murphy2013},
only few works focused on a simplification of these theories to scaling laws
that can be easily verified by the extraction of volume-integrated quantities from
multidimensional simulations. \citet{Murphy2013} performed a quantitative 
analysis of the interdependence of neutrino heating and non-radial instabilities 
with respect to the effect of turbulent motions on the average shock radius. 
In \citet{Mueller2015}, semi-empirical scaling laws were formulated that describe the relations 
between the turbulent kinetic energy and Mach number, shock deformation, and neutrino heating. 
In the following, guided by the results of \citet{Mueller2015}, we aim at
investigating to what extent the additional consideration of turbulent stresses in the 
gain layer can lead to a generalizable description of the explosion conditions,
being commonly applicable to all 18 simulations.

In order to derive the critical luminosity, we start with the spherical symmetric case, considering the scaling relations for
$\tau_\mathrm{adv}$,
\begin{equation}
\tau_\mathrm{adv} \propto \frac{R_\mathrm{s}^{3/2}}{\sqrt{M_\mathrm{NS}}},
\end{equation}
and $\tau_\mathrm{heat}$,
\begin{equation}
\tau_\mathrm{heat} \propto \frac{\left|\bar{e}_\mathrm{tot,g}\right|R_\mathrm{g}^2}{L_\nu \langle E_\nu^2\rangle}
\end{equation}
\citep[see][]{Janka2012a}. Here, $\bar{e}_\mathrm{tot,g}$ is the average 
mass-specific binding energy in the gain layer:
\begin{equation}
\bar{e}_\mathrm{tot,g} = \frac{E_\mathrm{tot,g}}{M_\mathrm{g}},
\end{equation}
where $E_\mathrm{tot,g}$ is defined in Eq.\,(\ref{eq:ene}).
$L_\nu$ is defined as the total
luminosity $L_\nu=L_{\nu_\mathrm{e}}+L_{\bar{\nu}_\mathrm{e}}$ of $\nu_\mathrm{e}$
and $\bar{\nu}_\mathrm{e}$, and $\langle E_\nu^2\rangle$ denotes
the weighted average of the mean squared energies of electron neutrinos and antineutrinos:
\begin{equation}
\langle E_\nu^2\rangle = \frac{L_{\nu_\mathrm{e}} \langle E_{\nu_\mathrm{e}}^2\rangle + L_{\bar{\nu}_\mathrm{e}} \langle E_{\bar{\nu}_\mathrm{e}}^2\rangle}{L_{\nu_\mathrm{e}}+L_{\bar{\nu}_\mathrm{e}}}.
\end{equation}
As in Eq.\,(\ref{eq:qheat}), the mean squared energies are defined as 
$\langle E_{\nu_\mathrm{e}}^2\rangle \coloneqq \langle \epsilon_{\nu_\mathrm{e}}^3\rangle / \langle \epsilon_{\nu_\mathrm{e}}\rangle$ 
and $\langle E_{\bar{\nu}_\mathrm{e}}^2\rangle \coloneqq \langle \epsilon_{\bar{\nu}_\mathrm{e}}^3\rangle / \langle \epsilon_{\bar{\nu}_\mathrm{e}}\rangle$.
According to \citet{Janka2012a}, the shock radius in spherical symmetry follows the relation
\begin{equation}
R_\mathrm{s} \propto \frac{\left(L_\nu \langle E_\nu^2\rangle\right)^{4/9}R_\mathrm{g}^{16/9}}{\dot{M}^{2/3}M_\mathrm{NS}^{1/3}}.
\label{eq:r_shock}
\end{equation}
By the use of these approximate scaling relations, the time-scale
criterion $\tau_\mathrm{adv}/\tau_\mathrm{heat}\sim1$ can be translated into
a critical luminosity condition which depends on the mass-accretion rate, the proto-neutron
star mass, the gain radius, and the average specific binding energy in the gain layer:
\begin{equation}
\left(L_\nu \langle E_\nu^2\rangle\right)_\mathrm{crit}\propto\left(\dot{M}M_\mathrm{NS}\right)^{3/5}\left|\bar{e}_\mathrm{tot,g}\right|^{3/5}R_\mathrm{g}^{-2/5}.
\end{equation}
Note that we do not omit  $\bar{e}_\mathrm{tot,g}$ and $R_\mathrm{g}$ in this relation.

For the multidimensional case, we follow \citet{Mueller2015} and consider the turbulent stresses of multidimensional flows in the gain layer by 
introducing an additional isotropic pressure contribution $P_\mathrm{turb}\approx\langle\delta v^2\rangle\rho\approx4/3\langle\mathrm{Ma}^2\rangle P$.
The consideration of this additional post-shock pressure leads to an increased advection time scale because of a 
larger radius of the stalled shock compared to Eq.\,(\ref{eq:r_shock}):
\begin{equation}
R_\mathrm{s} \propto \frac{\left(L_\nu \langle E_\nu^2\rangle\right)^{4/9}R_\mathrm{g}^{16/9}}{\dot{M}^{2/3}M_\mathrm{NS}^{1/3}}\left(1+\frac{4\langle\mathrm{Ma}^2\rangle}{3}\right)^{2/3}
\label{eq:r_shock_turb}
\end{equation}
\citep[see Appendix B of][]{Mueller2015}. Taking this modification into account, the scaling relation
for the critical luminosity now reads
\begin{equation}
\left(L_\nu \langle E_\nu^2\rangle\right)_\mathrm{crit} \propto\left(\dot{M}M_\mathrm{NS}\right)^{3/5}\xi_\mathrm{g},
\end{equation}
where the time dependent quantity $\xi_\mathrm{g}$ subsumes all gain-layer related properties:
\begin{equation}
\xi_\mathrm{g}\coloneqq\left|\bar{e}_\mathrm{tot,g}\right|^{3/5}R_\mathrm{g}^{-2/5}\left(1+\frac{4\langle\mathrm{Ma}^2\rangle}{3}\right)^{-3/5}.
\label{eq:xi}
\end{equation}
$\xi_\mathrm{g}$ can be used to correct 
$L_\nu \langle E_\nu^2\rangle$ with respect to the time-dependent
evolution of gain radius, binding energy, and turbulent pressure in the
gain layer, which lead to a time and model dependence of the critical luminosity condition for an explosion in addition
to its dependence on $M_\mathrm{NS}$ and $\dot{M}$:
\begin{equation}
\left(L_\nu \langle E_\nu^2\rangle\right)_\mathrm{crit,corr}\coloneqq\frac{1}{\xi_\mathrm{g}/\xi_\mathrm{g}^*}\left(L_\nu \langle E_\nu^2\rangle\right)_\mathrm{crit}.
\label{eq:lnu_crit}
\end{equation}
In order to obtain a meaningful comparison between different models, we 
also introduce a constant normalization factor $\xi_\mathrm{g}^*$ such that the correction is applied relative
to a reference model. This reference model can be chosen arbitrarily. For our analysis
we selected model s16.8 and evaluated $\xi_\mathrm{g}^*$ at the time when the 
ratio $\tau_\mathrm{adv}/\tau_\mathrm{heat}$ reaches unity:
\begin{equation}
\xi_\mathrm{g}^*\coloneqq \left.\left|\bar{e}_\mathrm{tot,g}\right|^{3/5}R_\mathrm{g}^{-2/5}\left(1+\frac{4\langle\mathrm{Ma}^2\rangle}{3}\right)^{-3/5}\right|^{\mathrm{s16.8}}_{\tau_\mathrm{adv}/\tau_\mathrm{heat}=1}.
\label{eq:xi_g}
\end{equation}
This, finally, leads to a generalized version of the critical condition, now applying to
the corrected values of $L_\nu \langle E_\nu^2\rangle$:
\begin{equation}
\left(L_\nu \langle E_\nu^2\rangle\right)_\mathrm{crit,corr}\propto\left(\dot{M}M_\mathrm{NS}\right)^{3/5}.
\label{crit_cond}
\end{equation}

The results of our analysis are shown in Fig.\,\ref{corr}, the corresponding correction factors
are given in Table\,\ref{tab2}. In addition to the time evolution
of the corrected and normalized values of 
$\left(L_\nu \langle E_\nu^2\rangle\right)_\mathrm{corr}=\left(\xi_\mathrm{g}/\xi_\mathrm{g}^*\right)^{-1}\left(L_\nu \langle E_\nu^2\rangle\right)$ 
versus $\left(\dot{M}M_\mathrm{NS}\right)^{3/5}$ (lower panel), we depict the instants 
when the ratio $\tau_\mathrm{adv}/\tau_\mathrm{heat}$ exceeds unity. In the 
upper panel, these points are shown with corrections (filled symbols) and without corrections 
(empty symbols). The success of the correction procedure is evident:
accounting for the additional dependence of the critical luminosity condition
on $\bar{e}_\mathrm{tot,g}$, $R_\mathrm{g}$, and in particular on the turbulent stresses
of multidimensional flows in the gain layer leads to the expected strong correlation
with $\dot{M}M_\mathrm{NS}$, and a generalized
critical curve (indicated by the black dashed line) appears which is valid for all 18 explosion models. Note that
the critical curve shows up as a straight line in Fig.\,\ref{corr} since we plot $\left(\dot{M}M_\mathrm{NS}\right)^{3/5}$
on the abscissa. All models approach the critical curve from the right and move upwards
after reaching the critical condition. The upward bending of the evolutionary tracks at the onset of explosion is caused by
a steep drop of $\xi_\mathrm{g}$ in the denominator while $\left(L_\nu \langle E_\nu^2\rangle\right)$ in the numerator evolves 
slowly. The decline of $\xi_\mathrm{g}$ occurs because an increase of $\left(1+4/3\langle\mathrm{Ma}^2\rangle\right)$ supports
the outward acceleration of the shock and, as a consequence, the specific binding 
energy of the gain layer, $\left|\bar{e}_\mathrm{tot,g}\right|$, plummets in addition. Interestingly, also the behavior of the models 
exploding at rather late times after bounce is correctly captured by this condition. This 
further underlines the general validity of the critical curve defined above. 

In view of the analysis of \citet{Radice2016}, which demonstrates that aside of the 
turbulent pressure other effects of turbulence, e.g. a term associated with
centrifugal support, play an equally important role, it is quite astonishing that
a simple correction by the turbulent pressure term in the critical luminosity condition 
seems to capture the overall effects of multidimensional fluid motions in the
gain layer remarkably well.

For comparison, we also show the trajectory of a model
from \citet{Heger2005} (m15b6\footnote{\url{http://www.2sn.org/stellarevolution/magnet/}}, 
simulated in axisymmetry (2D) without the consideration of rotational effects)
that does not explode. As indicated by the solid black line in the lower panel of Fig.\,\ref{corr},
this model does not reach the critical luminosity condition, but evolves in parallel to the critical curve
in downward direction. The fact that this model does not fulfill the necessary condition for 
a successful runaway is correctly mirrored by its time evolution in the 
$\left(L_\nu\langle E_\nu^2\rangle\right)_\mathrm{corr}-\left(\dot{M}M_\mathrm{NS}\right)^{3/5}$ plane
(Eqs.\,\ref{eq:xi}-\ref{crit_cond}).
In summary, the critical curve constructed as described above 
proves to be an excellent yardstick for the onset of the explosion and defines a reliable,
general criterion for the development of runaway conditions in the simulations.

\section{Conclusions}\label{concl}

Our study of 18 pre-supernova models in a range of 11 to 28 solar masses,
using 2D simulations with three-flavor, energy dependent, ray-by-ray-plus 
neutrino transport including the full set of state-of-the-art neutrino reactions and 
microphysics, underlines the viability of the neutrino-driven mechanism in 
axisymmetry. All investigated models explode and a systematic comparison of the 
model set shows that the explosions are strongly influenced by the 
pre-collapse structure of the progenitor star. 

If the progenitor exhibits a pronounced decline of the density at the Si/Si-O 
composition shell interface, the rapid drop of the mass-accretion rate at 
the time when the interface arrives at the shock front induces a steep reduction
of the accretion ram pressure. This causes a strong shock expansion supported by neutrino 
heating and thus favors an early explosion. Such a behavior is particularly
likely when the mass-accretion rate is high before the Si/Si-O interface passes
the shock. In this case the neutron star mass grows quickly and a high accretion
luminosity ensures a high neutrino heating rate even after the composition-shell
interface has fallen through the shock. If the progenitor structure does not 
exhibit a pronounced density jump at the Si/Si-O interface and the mass-accretion 
rate decreases more slowly, the models tend to explode rather late
when the mass-accretion rate has declined enough for the 
neutrino heating to overcome the accretion ram pressure. 

Due to initially rather short advection time scales, our simulations provide 
favorable conditions for the efficient growth of the SASI. Large-scale mass 
motions in the post-shock layer associated with low-mode oscillations of the 
supernova shock front along the symmetry axis mirror the vivid SASI activity in 
our models, and the final shock expansion is initiated by the growth of large 
bubbles supported by this instability. But also the strong influence of 
convection is visible: When the time-scale ratio approaches unity, the $\chi$
parameter increases above the critical value of 3. A comparison to the SASI and 
convection dominated models discussed by \citet{Fernandez2014} confirms 
the typical fingerprints of both convection and the SASI in our models, since the turbulent energy
spectra of our simulations show the characteristic
SASI peak at a spherical harmonics mode of $l=2$ as well as enhanced 
convective power at higher modes of $l=5-10$.

\capstartfalse
\begin{deluxetable}{cccc}
\tabletypesize{\scriptsize}
\tablecaption{Correction factors for the critical luminosity \\used in Eq.\,(\ref{eq:lnu_crit})\label{tab2}}
\tablehead{
\colhead{model} & \colhead{$(\xi_\mathrm{g}/\xi_\mathrm{g}^*)^{-1}$} & \colhead{model} & \colhead{$(\xi_\mathrm{g}/\xi_\mathrm{g}^*)^{-1}$}}
\startdata
\textit{Model Set I} & & \textit{Model Set II} \\
s12-2007 & 0.86 & s11.2 & 1.25 \\
s15-2007 & 0.83 & s12.4 & 0.92 \\
s20-2007 & 1.07 & s13.2 & 0.86 \\
s25-2007 & 0.95 & s14.4 & 0.75 \\
& & s16.8 & 1.00 \\
& & s17.2 & 1.08 \\
& & s18.4 & 0.80 \\
& & s19.6 & 0.98 \\
& & s20.2 & 1.04 \\
& & s21.6 & 0.95 \\
& & s22.4 & 0.94 \\
& & s26.6 & 1.05 \\
& & s27.0 & 1.01 \\
& & s28.0 & 1.00 
\enddata
\tablecomments{$\xi_\mathrm{g}$ and $\xi_\mathrm{g}^*$ are defined in 
Eqs.\,(\ref{eq:xi}) and (\ref{eq:xi_g}).}
\end{deluxetable}
\capstarttrue

The investigation of a larger set of self-consistent CCSN simulations naturally
leads to the question of common properties shared by all models that govern the onset
of the successful explosions. Although the time-scale criterion proves to be a reliable 
diagnostic parameter for runaway, obvious correlations with specific 
values of other variables discussed in Sect.\,\ref{res_dis}
cannot be found. Following the approach suggested by \citet{Mueller2015} to account for 
the role of non-radial instabilities in the concept of a critical neutrino luminosity
for the onset of neutrino-driven explosions, we generalize the critical luminosity 
relation by including corrections for the effects of turbulent stresses (and of other
time-dependent parameters) in the gain layer (see Eqs.\,\ref{eq:xi}-\ref{crit_cond}). 
This relation defines a direct proportionality
between the corrected product of $\left(L_\nu\langle E_\nu^2\rangle\right)_\mathrm{corr}$ and 
$\left(\dot{M}M_\mathrm{NS}\right)^{3/5}$ and captures the explosion behavior of all 18 models
in an excellent way, thus reliably determining the conditions necessary for the onset of the
runaway. Our $\left(L_\nu\langle E_\nu^2\rangle\right)_\mathrm{corr} - \dot{M}M_\mathrm{NS}$
relation (see Eq.\,(\ref{crit_cond}) and Fig.\,\ref{corr}) 
leads to a considerable reduction of the scattering of the
critical runaway condition of all models compared to the uncorrected case as well as compared
to the $L_\nu - \dot{M}$ condition discussed by \citet{Suwa2016}, see Figure 18 there.

Since recent 2D core-collapse simulations by \citet{Bruenn2013,Bruenn2016}, \citet{Dolence2015},
\citet{OConnor2015}, and \citet{Skinner2015} focused on four progenitors models of 
\citet{Woosley2007} that are extensively investigated also in this work, 
detailed comparisons between different codes applied to the CCSN 
problem become possible now. At first glance, the differences between the 
results give reasons for concern \citep[for a cautious effort of a comparative discussion see also][]{Janka2016}: the same progenitor models fail to explode 
\citep[e.g.][but with Newtonian gravity and different EoS]{Dolence2015}, explode very early at a time that is nearly independent of the 
progenitor mass \citep[e.g.][]{Bruenn2013,Bruenn2016}, or explode later, showing 
a strong influence of the respective progenitor structure (this work). However, \citet{OConnor2015} demonstrated
that Newtonian gravity \citep[as applied by][]{Dolence2015} is not favorable for explosions while a 
relativistic potential is. 
\citet{Skinner2015} reported differences in the dynamical evolution of the four progenitor models with M1 and 
ray-by-ray neutrino transport, the latter favoring explosions. But these results are in conflict with the M1 models of \citet{OConnor2015}, 
which show overall agreement with the ray-by-ray-plus results presented in our work. Curiously, the 
differences observed by \citet{Skinner2015} decreased when the resolution of their simulations was enhanced.
Good overall agreement with our results was also demonstrated 
in a recent conference talk at FOE2015\footnote{\url{https://www.physics.ncsu.edu/FOE2015/PRESENTATIONS/FOE2015_kotake.pdf}}
by K. Kotake, who presented his simulations for a subset of cases of our Model Set II.

A profound analysis of similarities and differences of simulations depending on the applied 
codes and microphysics is demanded to shed light on the sensitivity of the CCSN dynamics to the 
approximations still used in current simulations. Particular attention will have to be paid
to the possible role of code- and method inherent numerical perturbations, which might foster 
the growth of post-shock instabilities and could have important consequences for the onset
of explosions \citep{Couch2013a,Couch2015,Mueller2015}. A close comparison will help 
putting present CCSN simulations on a touchstone and will point to 
necessary improvements in the modeling of this important astrophysical
problem.

\acknowledgments

We thank R. Bollig and T. Ertl for helpful discussions and support and the 
anonymous referee for interesting comments. 
The project was funded by the Deutsche Forschungsgemeinschaft through 
grant EXC 153 and by the European Research Council through 
grant ERC-AdG No.\ 341157-COCO2CASA. B.M. received support by the 
Australian Research Council through a
Discovery Early Career Researcher Award (grant DE150101145). 
We acknowledge computing time from
the European PRACE initiative on SuperMUC (GCS@LRZ, Germany), 
Curie (GENCI@CEA, France), and MareNostrum (BSC-CNS, Spain).
Postprocessing was done on Hydra of the Max Planck Computing and Data Facility.

\appendix

\begin{figure*}
\includegraphics[width=\textwidth]{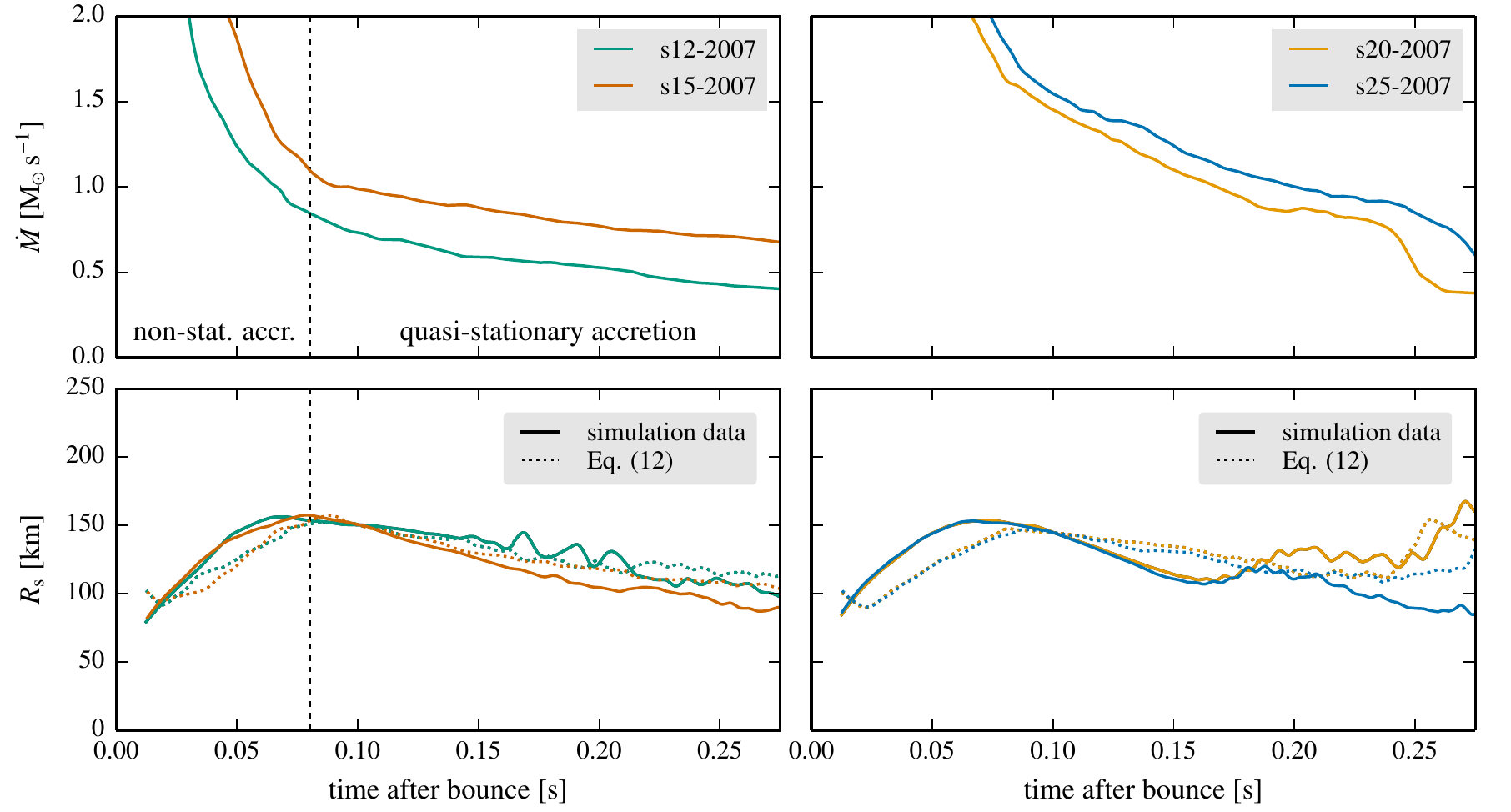}
\caption{Mass-accretion rates (top panels) and mean shock trajectories (bottom panels) for the models of Set I with the lower-mass 
cases on the left and the higher-mass cases on the right. In the lower panels, solid lines denote the simulation data, while
the dotted lines provide the analytic approximation of Eq.\,(\ref{eq:rs}). The latter holds for quasi-stationary accretion 
conditions, which apply better in the low-mass cases after an initial phase of high mass-accretion rates. In the high-mass models,
the accretion rates continue to remain on a high level with a steep decline for a longer period of time, for which reason  
the analytic approximation roughly captures the general trend but does not show the good quality of the quantitative agreement visible
in the lower left panel.}
\label{shock_radii}
\end{figure*}

The following appendices provide information on several aspects of discussion in detail.
First, we demonstrate the viability of Eq.\,(\ref{eq:rs}) for a rough description of the steady-state
evolution of the radius of the accretion shock. Second, we follow \citet{Nakamura2015} and present 
correlations of some explosion properties with the compactness parameter $\xi_{2.0}$ defined by Eq.\,(\ref{eq:comp}).

Moreover, we aim at studying the resolution and stochasticity dependence 
of our results with respect to the point in time when the explosion sets in. This 
is intended to further validate the connection between progenitor structure and
post-bounce evolution that is evident in our simulations and has been extensively
discussed in this paper. We also vary the chosen transition density between
the high- and the low-density EoS and study the influence of different treatments
of energy conservation on the simulation outcome.

In addition, we will test the effects of differences in the 
employed neutrino physics compared to the models of \citet{Bruenn2013,Bruenn2016}
on the shock and neutron star radii. Even in 1D, these quantities
differ significantly between the simulation results of \textsc{Prometheus-Vertex} and 
the published results of the \textsc{Chimera} code used by the Oak Ridge group. We note, however, that the Oak Ridge
group has recently presented 1D results for ``Series C'' models \citep{Lentz2015}, where the shock
radii are considerably smaller than in the previous ``Series B'' 1D models of 
\citet{Bruenn2013,Bruenn2016}, and therefore closer to our results obtained with
\textsc{Prometheus-Vertex}.

We emphasize that our multidimensional code retains spherical symmetry exactly if no seed perturbations 
are applied. Despite their potentially important role for the development of post-shock instabilities
\citep{Couch2013a,Couch2015,Mueller2015}, we have not varied the recipe of random seeds
employed in this study but have constrained ourselves to the seeding method described in
Sect.\,\ref{num_set} for all models.\\

\section{Evolution of mean shock radius and analytic approximation}\label{evo_shock}

In order to demonstrate the viability of Eq.\,(\ref{eq:rs}) for a rough description of the time evolution of the mean shock radius,
Fig.\,\ref{shock_radii} displays the shock trajectories for the four models of our Set I. 
Both the simulation data (solid lines) and the proportionality
relation according to Eq.\,(\ref{eq:rs}) (dotted lines; the normalization constant of this relation is 
chosen such that the relation matches the simulation data at 0.1\,s) are shown. 
Eq.\,(\ref{eq:rs}) describes the simulation data very closely only in a 
time interval in which steady-state conditions are roughly fulfilled.
This is the case after the early maximum of the shock expansion (the initial shock expansion is 
driven by the non-stationary accumulation of an accretion mantle around the neutron star) and 
before the development of strong non-radial mass motions in the post-shock flow.
A steep decline of the mass-accretion rate continues for a longer period of time in 
the two more massive models, while the two less massive cases reach a quasi-stationary accretion 
state after about 80\,ms of post-bounce evolution (see top row of Fig.\,\ref{shock_radii}). Therefore
the requirement of stationarity is better fulfilled for the two less massive stars, for which 
reason the proportionality relation of Eq.\,(\ref{eq:rs}) agrees better with the simulation data.
Since the low-mass cases explode only late, their early conditions are farther away from the
threshold to explosion and multidimensional effects play a minor role, whereas such effects are slightly
more visible in the two more massive models.

\section{Explosion properties and compactness parameter}

\begin{figure}
\includegraphics[width=\columnwidth]{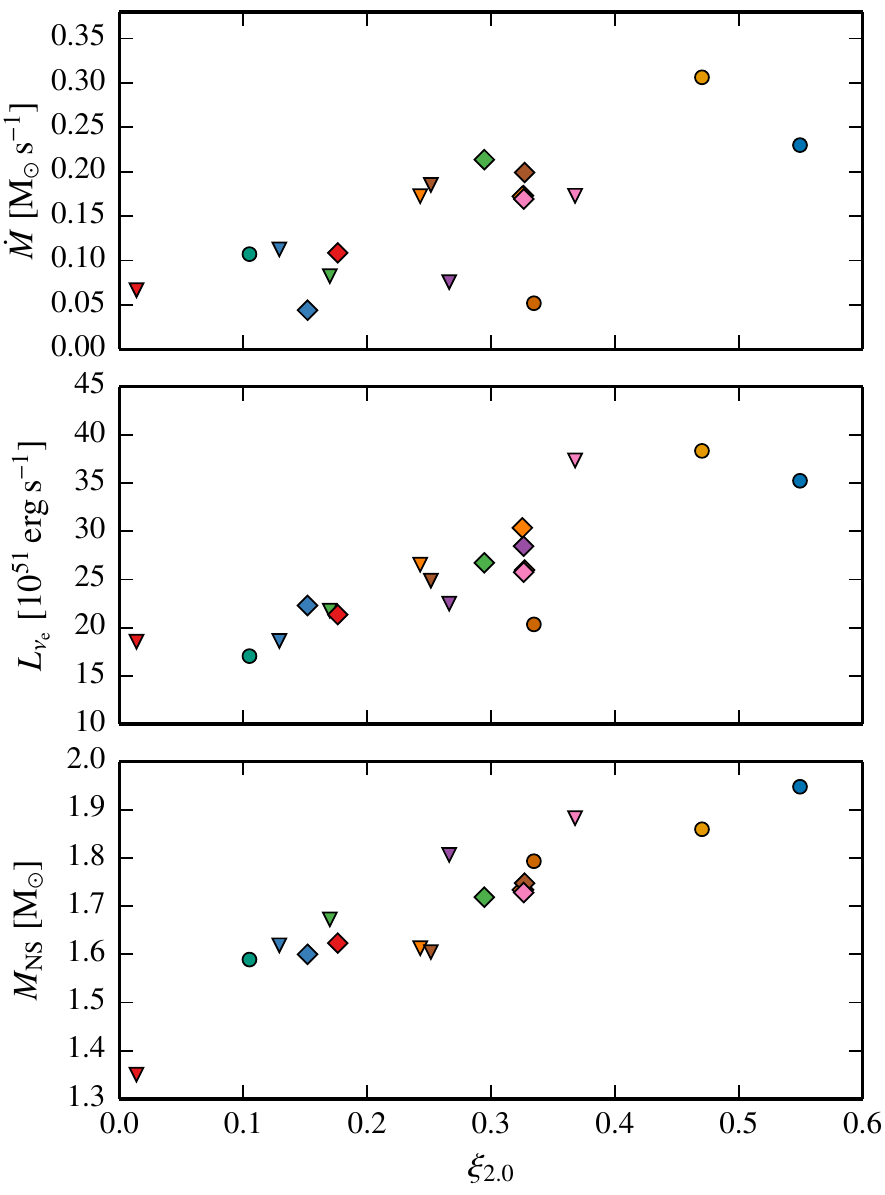}
\caption{Mass-accretion rate, electron neutrino luminosity, and proto-neutron star mass as
functions of the compactness parameter $\xi_{2.0}$ (from top to bottom). Following \citet{Nakamura2015}, the proto-neutron star
mass is given at the final time of the simulation, the two other quantities are evaluated at the time
when the mean shock radius reaches 400\,km. Circles denote the models of Set I, and triangles (diamonds) indicate the models of Set II with lower (higher) ZAMS 
masses in line with the classification and color coding used in previous figures.}\label{chi2.0}
\end{figure}

In Fig.\,\ref{chi2.0}, the mass-accretion rate, $\dot{M}$, the electron neutrino luminosity, $L_{\nu_\mathrm{e}}$, and 
the mass of the proto-neutron star, $M_\mathrm{NS}$, are shown for our 18 explosion models as functions of the compactness parameter
$\xi_{2.0}$ (cf.\ Eq.\,\ref{eq:comp}). As in Table\,1, the compactness parameter is 
calculated from the pre-supernova model (which in this case is identical to the value at bounce). Following \citet{Nakamura2015},
$\dot{M}$ and $L_{\nu_\mathrm{e}}$ (as defined in Sect.\,\ref{general_properties}) are evaluated at the time when the mean shock radius
reaches a value of 400\,km, while $M_\mathrm{NS}$ is given at the final time of our simulations.
We constrain the cases of Fig.\,\ref{chi2.0} to a single value of $\xi_M$ \citep[different from][]{Nakamura2015}, since choices
of $1.5\lesssim M \lesssim 2.5$ show similar correlations.
Although our model set exhibits the same increasing trends found by \citet{Nakamura2015},
only 18 data points do not provide sufficient statistics for a meaningful derivation 
of correlations. The observed trends can also be expected for fundamental physical reasons 
and are therefore not astonishing: For models with a higher compactness
parameter $\xi_{2.0}$, the mass coordinate of $2.0\,\mathrm{M}_\odot$ is located at a smaller radius $R(M=2.0\,\mathrm{M}_\odot)$ 
than for models with lower compactness. The same mass being compressed into a smaller sphere of radius $R(M=2.0\,\mathrm{M}_\odot)$ then 
translates into a longer-lasting high mass-accretion rate (Fig.\,\ref{chi2.0}, top panel), leading to a higher 
accretion luminosity (Fig.\,\ref{chi2.0}, middle panel) and to a higher proto-neutron star mass (Fig.\,\ref{chi2.0}, bottom panel). 
But we would also like to underline that, despite of the overall rough trends, the significant scatter
of the depicted quantities points towards peculiar model characteristics which cannot be captured sufficiently well by 
a single parameter like the compactness. As discussed and demonstrated in Sect.\,\ref{Crit}, the formulation of a criterion that 
reliably determines the development of runaway conditions in multidimensional simulations especially requires 
a proper consideration of the model-dependent effects of non-radial mass motions.

\section{Resolution Dependence and Stochasticity of the Results}\label{res_dep}

\begin{figure}
\includegraphics[width=\columnwidth]{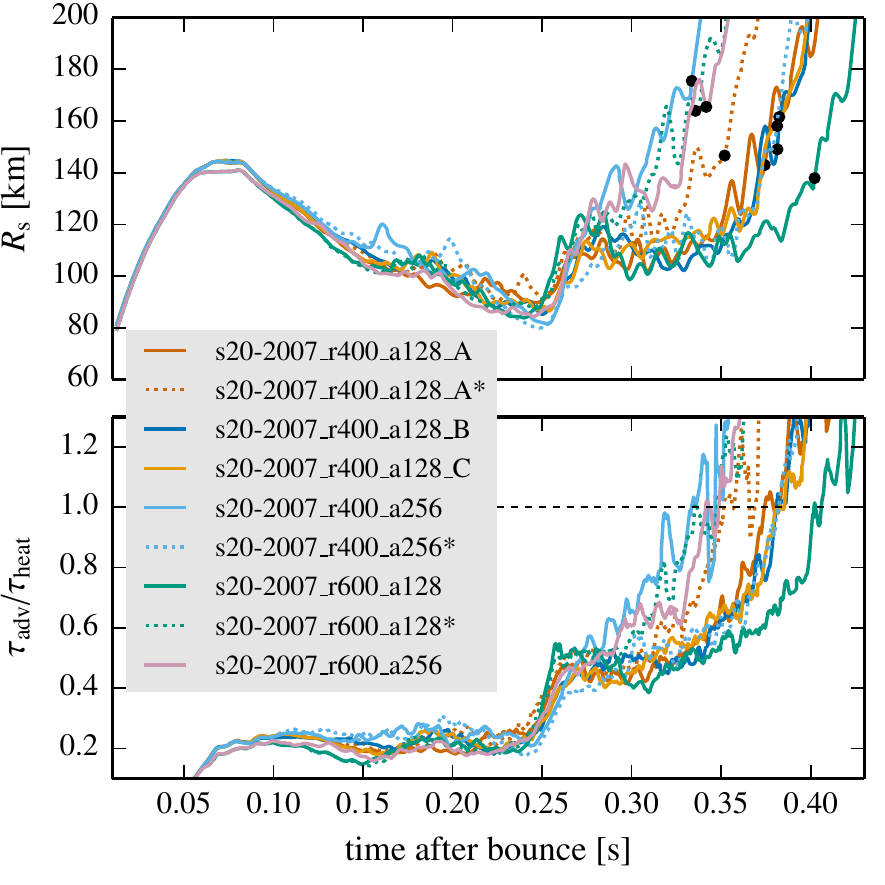}
\caption{Time evolution of shock radius (upper panel) and time-scale ratio
(lower panel) for a set of nine 2D simulations for the same progenitor model, 
but with different angular and radial resolutions and with different random
perturbations for seeding non-radial hydrodynamic instabilities. The black dots in the top panel
mark the point in time when the ratio $\tau_\mathrm{adv}/\tau_\mathrm{heat}$
reaches unity. All quantities are angle-averaged and the curves are smoothed by running averages of
5\,ms.\label{res_test}}
\end{figure}

For the resolution study, we chose model s20-2007 of 
\citet{Woosley2007}. The setups of the simulations are listed in Table\,
\ref{tab3}. Besides two different angular resolutions of 128 and 256 angular 
zones, radial grids of initially 400 and 600 zones (both gradually further refined 
during the simulations) were used, and various combinations of the highly and 
moderately resolved angular and radial grids were tested. We also varied the  
random seeds (but without changing the seeding recipe) 
for the density perturbations introduced 10\,ms after bounce (compare model 
s20-2007\_r400\_a128\_A, s20-2007\_r400\_a128\_B, and s20-2007\_r400\_a128\_C). 
This affects only the perturbation 
pattern, the perturbation amplitude of 0.1\,\% in density was the same for all models. 
In order to test for the stochasticity of the results, the models with names 
appended by an asterisk are just a repetition of the simulations without asterisk
for the same initial conditions (i.e., also the same
perturbations). 

The numerical setup of the models was identical to the description in 
Sect.\,\ref{num_set} except for several code improvements that were only used
in the simulations of this section. Besides minor changes 
this includes a more sophisticated treatment of total energy 
conservation \citep[cf.][]{Mueller2010} and the correction of an erroneously applied 
identity of the charged-current neutrino absorption coefficient in eight-cell OpenMP patches.
While the latter improvement has no noticeable effects on the results of test calculations,
the improved treatment of the total energy conservation leads to slightly smaller shock 
radii at earlier times ($\Delta R_\mathrm{s}\lesssim 10\,\mathrm{km}$ at the time of
maximal shock expansion, $R_\mathrm{s,max}\sim150\,\mathrm{km}$),
and we observe a somewhat delayed ($\sim 70\,\mathrm{ms}-100\,\mathrm{ms}$) 
development of a runaway situation compared to the s20-2007 case presented in Sect.\,\ref{res_dis}
(compare Figs.\,\ref{res_test} and \ref{gen_prop}). 
But as we will show in the following, stochasticity seems to be the key determinant
for the exact timing of the onset of explosion.

Since the hydrodynamic flow behind the shock front evolves highly non-linearly and 
in a chaotic way, differences in the detailed post-bounce dynamics of the presented simulations are 
expected, even if initial conditions and grid resolutions are identical. After 
150\,ms, this can be observed in the evolution of the shock radius and the time-scale 
ratio of models s20-2007\_r400\_a128\_A and s20-2007\_r400\_a128\_A* 
shown in Fig.\,\ref{res_test}. The stochastic nature of the developing non-radial flow in the 
post-shock layer results in a difference of $\sim30\,\mathrm{ms}$ between the 
times when the critical condition $\tau_\mathrm{adv}/\tau_\mathrm{heat}\gtrsim1$ 
is reached (see Fig.\,\ref{res_test}). \new{Similar stochastic differences can be observed for the models with
higher resolution (compare the time evolution of models s20-2007\_r600\_a128 with s20-2007\_r600\_a128* 
or s20-2007\_r400\_a256 with s20-2007\_r400\_a256*). It is remarkable that not even 
in the case of the same initial conditions our simulations
completely agree in the details of their time evolution. This can be explained by the applied 
compiler optimizations which are chosen to enhance the code performance, but also
marginally influence the precision of floating-point operations\footnote{We confirmed by tests that running the simulations
without any compiler optimization allows us to reproduce results of simulation runs in an exact way, starting from
the same initial perturbation patterns.}. Further enhanced by the
turbulent, non-linear evolution of the hydrodynamic dynamics behind the shock, these minimal differences can
lead to a certain spread in the evolution of the models.}

\new{Models} with a higher angular resolution of 256 zones seem to show a 
trend towards a slightly \textit{earlier} runaway than the simulations with 128 angular 
zones. This is in accordance with the results of \citet{Hanke2012} 
and their set of simulations using only a simplified and parametrized neutrino 
treatment. However, the rather small difference in time when the critical condition is 
met compared to \new{models s20-2007\_r400\_a128\_A* and s20-2007\_r600\_a128*}, which show the earliest 
runaway of all models with lower angular resolution, again suggests stochastics 
as likely main reason for the observed differences between the models. This is also the 
case for the models with initially higher radial resolution: Only differences of 
the order of a few tens of milliseconds can be observed concerning the points in 
time when the runaway sets in.

\capstartfalse
\begin{deluxetable}{lcc}
\tabletypesize{\scriptsize}
\tablecaption{Tests of resolution and EoS treatment\label{tab3}}
\tablehead{
\colhead{model} & \colhead{\# of radial zones} & \colhead{\# of angular zones}}
\startdata
\textit{Resolution tests:} \\
s20-2007\_r400\_a128\_A & 400 & 128 \\
s20-2007\_r400\_a128\_A* & 400 & 128 \\
s20-2007\_r400\_a128\_B & 400 & 128 \\
s20-2007\_r400\_a128\_C & 400 & 128 \\
s20-2007\_r400\_a256 & 400 & 256 \\
s20-2007\_r400\_a256* & 400 & 256 \\
s20-2007\_r600\_a128 & 600 & 128 \\
s20-2007\_r600\_a128* & 600 & 128 \\
s20-2007\_r600\_a256 & 600 & 256 \\
\tableline \\[-5pt]
\textit{Transition density tests:} \\
s20-2007\_r400\_a88\_rho3.4e7 & 400 & 88 \\
s20-2007\_r400\_a88\_rho6.0e7 & 400 & 88 \\
s20-2007\_r400\_a88\_rho3.0e8 & 400 & 88
\enddata
\tablecomments{Naming convention: numerical values following the small letters `r' and `a' indicate the 
numbers of radial and angular grid cells, respectively. Capital letters A, B, and C at the end of the model name
denote different patterns of density perturbations imposed 10\,ms after bounce.
\new{Models with an asterisk are just a repetition of the models without asterisk
with the same initial conditions.} In the cases of the tests for the EoS transition density, 
the respective density values are also given. In all resolution tests, a 
transition density of $3.0\times10^8\mathrm{g\,cm}^{-3}$ was applied.}
\end{deluxetable}
\capstarttrue

In a similar resolution study performed by \citet{Hanke2014} for model s27.0,
higher angular resolution hardly made any difference for the simulation
results, whereas higher radial resolution (combined with high angular resolution)
led to a delay of the runaway. While our study also shows a delay in the case of
model s20-2007\_r600\_a128, where only the radial resolution was increased, our two 
models with enhanced angular and enhanced angular and radial resolution, respectively, explode 
rather early. \new{This is also the case for model s20-2007\_r600\_a128*.}
The fact that the resolution changes do not produce uniform results
again underlines the strong influence of stochastic turbulent motions on the onset
of explosion.

Overall, the results of our resolution study can be summarized as follows. 
Although the time difference between the simulation showing the earliest runaway 
and the simulation exploding latest amounts to $\sim80\,\mathrm{ms}$ (defined by 
the point in time when the critical time-scale ratio is reached), the development of the 
explosion is mainly triggered by the arrival of the Si/Si-O composition shell 
interface at the shock. When the ram pressure of the infalling material is 
reduced to such an extent that neutrino heating can revive the shock, the 
explosion is initiated. The associated development of turbulent motions due to 
convection and the SASI promotes the shock revival by increasing the dwell time 
of matter in the gain layer and additional support of the shock by turbulent momentum
flows and pressure, and the stochasticity of these fluid motions 
finally leads to the moderate variance regarding the further evolution of the 
explosion. Different patterns of initial density perturbations or an 
energy cascade better resolved by a higher number of grid zones affect the evolution of 
the turbulent motions, but according to our resolution study, these are only 
secondary effects. This underlines the validity of our conclusions regarding the 
strong connection between post-bounce evolution and pre-collapse structure as 
evident in the results presented in the main text of our paper.

\section{Influence of pre-bounce EoS transition density}

\begin{figure}
\includegraphics[width=\columnwidth]{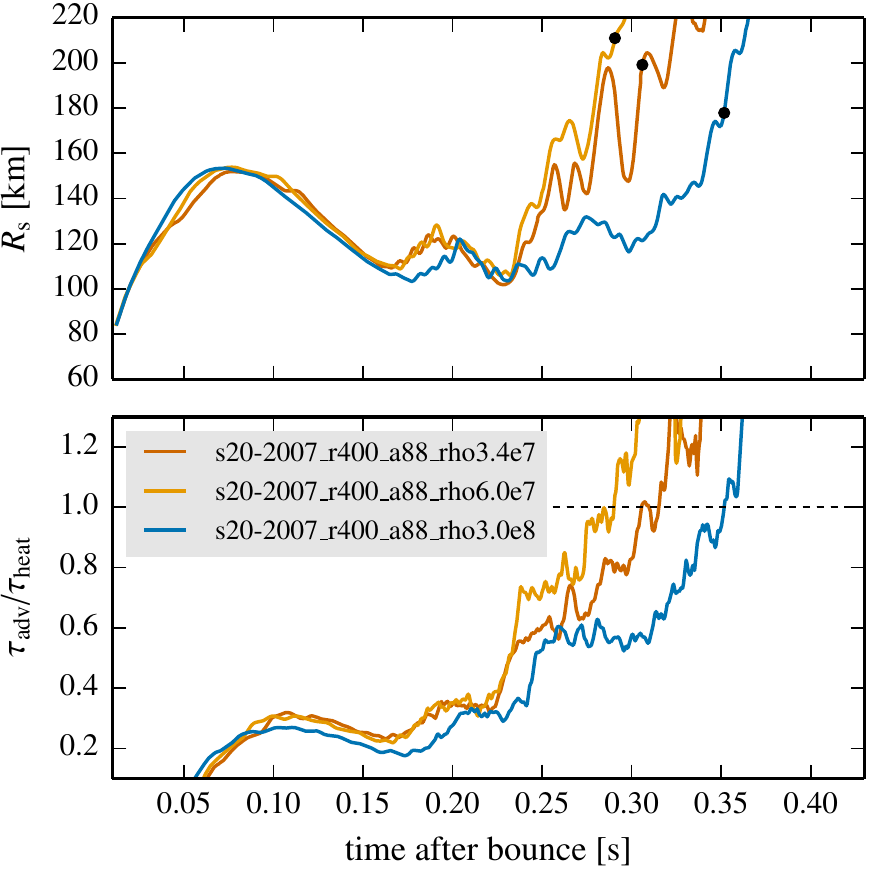}
\caption{Time evolution of shock radius (upper panel) and time-scale ratio
(lower panel) for a set of three 2D simulations for the same progenitor model, 
but with different choices for the transition density between the high- and the
low-density EoS before bounce. The black dots in the top panel
mark the point in time when the ratio $\tau_\mathrm{adv}/\tau_\mathrm{heat}$
reaches unity. All quantities are angle-averaged and the curves are smoothed by running averages of
5\,ms.\label{res_test2}}
\end{figure}

In addition to the tests of resolution and stochastic effects,
we also varied the choice of the density at which the transition from
the high-density to the low-density EoS is placed 
\citep[cf.][]{Rampp2002b,Buras2006b}. During the collapse phase until core bounce, the values for the 
transition density were chosen to be $3.4\times10^7\,\mathrm{g\,cm}^{-3}$, 
$6.0\times10^7\,\mathrm{g\,cm}^{-3}$, and $3.0\times10^8\,\mathrm{g\,cm}^{-3}$, respectively (see Table\,\ref{tab3} and Fig.\,\ref{res_test2}).
The last value was the one used in all 18 simulations presented in Sect.\,\ref{res_dis}. In 
the post-bounce phase the transition density was moved to $10^{11}\,\mathrm{g\,cm}^{-3}$ in all cases,
because below this density nucleon interactions play a negligible role and our low-density 
NSE solver allows for the consideration of a larger set of nuclear species during
the expansion phase of the high-entropy shock- and neutrino-heated gas, connecting
smoothly to the nuclear freeze-out and nuclear burning in the shock-accelerated ejecta.
For the three simulations, the same code version as in Sect.\,\ref{res_dis} was 
applied, and the number of angular zones was reduced to 88. Even though the time
when the critical time-scale ratio is reached differs by $\sim80\,\mathrm{ms}$ 
(see Fig.\,\ref{res_test2}) between the simulations, this
difference is still in the ballpark of the stochastic effects discussed above, 
showing that the exact choice of the transition density (in contrast to the progenitor
structure) has no major impact on the post-bounce dynamics beyond the level of variation
associated with stochastic fluctuations. The shift is a consequence of differences
in the infall profile of the outer Fe-core and Si-shell layers, which develop between the
start of the simulations and core bounce, i.e. before the time when the EoS transition
density is set to $10^{11}\,\mathrm{g\,cm}^{-3}$ in all cases.

\section{Neutrino-pair conversion and scattering processes and effective mass corrections}

\begin{figure}
\includegraphics[width=\columnwidth]{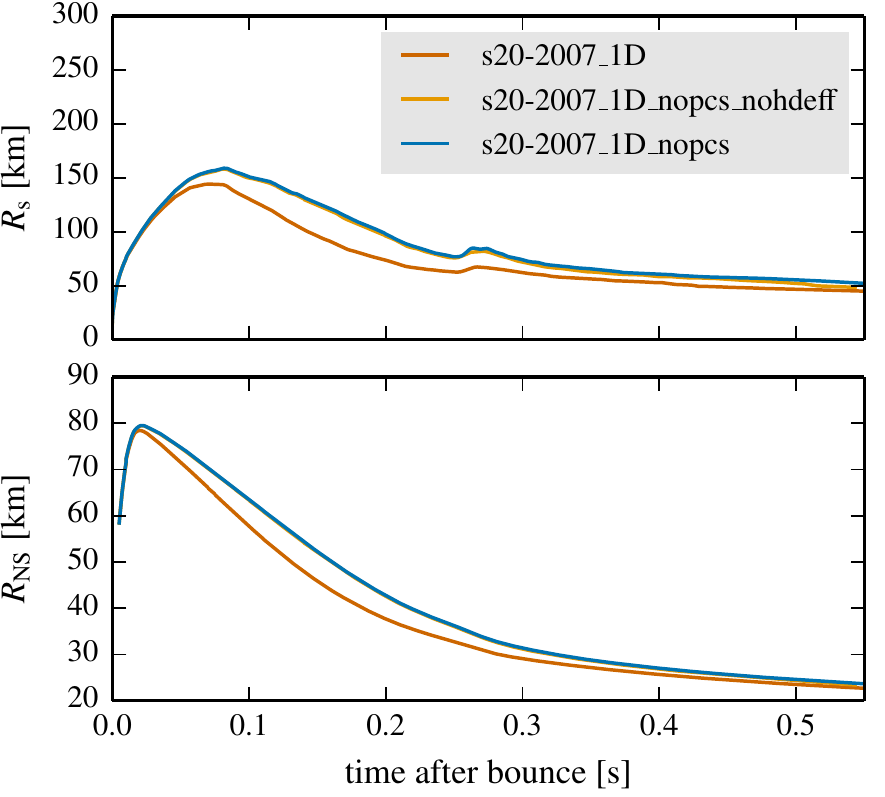}
\caption{Time evolution of shock radius (upper panel) and neutron star 
radius (lower panel) for three 1D simulations applying different neutrino physics.
The neutron star radius is defined by the density 
surface at $10^{11}\,\mathrm{g}\,\mathrm{cm}^{-3}$.
While model s20-2007\_1D includes the full set of neutrino reactions, pair-conversion
processes between different $\nu$ flavors and $\nu\nu$ scattering reactions are disabled in 
model s20-2007\_1D\_nopcs, and high-density nucleon
correlations as well as effective mass corrections are additionally switched off in model
s20-2007\_1D\_nopcs\_nohdeff. Note that the lines for models s20-2007\_1D\_nopcs and 
s20-2007\_1D\_nopcs\_nohdeff fall on top of each other. The curves are smoothed by running averages of
5\,ms.\label{rates_test}}
\end{figure}

According to \citet{Bruenn2013,Bruenn2016}, the employed neutrino physics in the 
\textsc{Chimera} and \textsc{Prometheus-Vertex} simulations is similar except for the 
treatment of in-medium nucleon correlations and nucleon-mass corrections at high densities, 
$\nu_\mathrm{e}\bar{\nu}_\mathrm{e}\leftrightarrow\nu_{\mu\tau}\bar{\nu}_{\mu\tau}$ pair-conversion
processes, and pure neutrino-scattering reactions \citep{Buras2003a},
which are not included in the \textsc{Chimera} code. In order to test if 
these differences in the neutrino physics can explain the considerable differences between the results
of the Oak Ridge group (``Series B'') and the Garching group concerning shock and neutron star radii,
we performed three 1D simulations of the 20\,M$_\odot$ progenitor of 
\citet{Woosley2007}. 

Besides a simulation with the full neutrino physics (s20-2007\_1D), a second 
simulation was run with neutrino pair-conversion and scattering processes switched off (s20-2007\_1D\_nopcs)
and a third simulation with high-density nucleon
correlations and in-medium mass corrections (s20-2007\_1D\_nopcs\_nohdeff) additionally disabled.
The improved treatment of energy conservation described in Appendix \ref{res_dep} was also applied here.
The results are shown in Fig.\,\ref{rates_test}. While the omission of neutrino pair-conversion and scattering processes
leads to slightly larger shock and neutron star radii, the high-density effects
do not have any significant additional influence on the displayed quantities. Therefore, the differences
in the applied neutrino physics cannot account for the larger shock
and neutron star radii observed even in the 1D simulations with the \textsc{Chimera} code \citep{Bruenn2013,Bruenn2016}.
Note that the neutron star radii depicted in Figure 3 of \citet{Bruenn2013} seem to be 
-- in contrast to our previous statement -- smaller than
those for our models shown in Fig.\,\ref{exp} (second panel from top). However, since
the proportionality between shock radius and neutron star radius given in Eq.\,(\ref{eq:rs})
should also hold for the results of \citet{Bruenn2013} and the smaller neutron star radius values
shown by \citet{Bruenn2013} do not make sense for their larger shock radii, we attribute this discrepancy to 
the accidental omission of a scaling factor in their figure.

Furthermore, a comparison to Figure\,4 of \citet{Steiner2013} shows good agreement
between the results of our simulations with \textsc{Prometheus-Vertex} and the 
1D simulations with the \textsc{Agile-Boltztran} code, 
an independently developed general relativistic hydrodynamics solver with 
three-flavor Boltzmann neutrino transport. In view of this unsatisfactory situation further comparisons
between different codes employed for CCSN simulations
are indispensable in order to determine the origin of the current discrepancies in the field.

\bibliographystyle{apj}
\bibliography{ccsne}

\end{document}